\documentclass{birkjour}
\usepackage{graphics}
 \theoremstyle{definition}
 
 \theoremstyle{remark}

 \numberwithin{equation}{section}

\begin{document}

%-------------------------------------------------------------------------
% editorial commands: to be inserted by the editorial office
%
%\firstpage{1} \volume{228} \Copyrightyear{2004} \DOI{003-0001}
%\seriesextraline{This is the Concrete Title of this Book\br H.E. R and S.T.C. W, Eds.}
%\firstpage{1}

\title[Renormalization Group and the Ricci Flow]
 {Renormalization Group and the Ricci Flow}
 
%----------Author 1
\author[M. Carfora]{Mauro Carfora}

\address{%
Dipartimento di Fisica Nucleare e Teorica\\
and Istituto Nazionale di Fisica Nucleare\\
Sezione di Pavia\\
Via Bassi 6\\
27100 Pavia\\
Italy}

\email{mauro.carfora@pv.infn.it}

%\thanks{This work was completed with the support of our \TeX-pert.}

\subjclass{Primary 99Z99; Secondary 00A00}

\keywords{Ricci flow, Quantum Field Theory}

\date{December 10, 2009}
%\dedicatory{}
%%% ----------------------------------------------------------------------

\begin{abstract}
We discuss from a geometric point of view the connection between the renormalization group flow for non--linear sigma models and the Ricci flow. This offers new perspectives  in providing a geometrical landscape for 2D quantum field theories. In particular we argue that the structure of  Ricci flow singularities suggests a natural way for extending, beyond the weak coupling regime, the embedding of the Ricci flow into the renormalization group flow. 
\end{abstract}

%%% ----------------------------------------------------------------------
\maketitle
%%% ----------------------------------------------------------------------
%\tableofcontents

\section{Introduction}
Among the many significant ideas and developments that connect Mathematics with contemporary Physics one that would have certainly intrigued Riemann himself is the role that Quantum Field Theory (QFT) plays in Geometry and Topology. We can argue back and forth on the relevance of such a role, but the perspective QFT offers  is often surprising and far reaching. Examples abound, and a fine selection is provided by the revealing insights offered by Yang--Mills theory into the topology of $4$--manifolds, by the relation between Knot Theory and topological QFT, and most recently by the interaction between Strings, Riemann moduli space, and enumerative geometry. Doubtless many of the most striking connections suggested by physicists failed to pass the censorship of the Department of Mathematics, and so do not appear in the above official list. However, the role of these connections does not wholly  fade out as a source of inspiration of mathematically interesting results. As an example that I would like to enter in the above inventory, and which passed  into rather general use, is the role of path integral techniques and of the attendant renormalization group philosophy. As ill--defined these techniques may be, if we give them some degree of mathematical acceptance then the geometrical perspective they afford is always quite non--trivial and extremely rich. It is  within such a framework that we shall examine here some aspects of the relation between the renormalization group analysis of a particular class of QFTs and the Ricci flow. In recent years, Ricci flow has been the point of departure and the motivating example for important developments in geometric analysis, most spectacularly for G. Perelman's proof \cite{18,19,20}
of Thurston's geometrization program \cite{thurston1,thurston2} for three-manifolds and of the attendant Poincar\'e conjecture. For one of those strange circumstances  not unusual in the history of Science, the Ricci flow, introduced in the early '80s by    Richard Hamilton\cite{11},  independently appeared on the scene also in Physics. Indeed, Daniel Friedan, studying the weak coupling limit of the renormalization group flow for non-linear sigma models, introduced \cite{DanThesis,DanPRL, DanAnnPhys} what later on came to be known as the Hamilton--DeTurck version of the Ricci flow. This QFT avatar of the Ricci flow was largely ignored in geometry until G. Perelman acknowledged \cite{18} that in his groundbreaking analysis he was somewhat inspired by the role that the effective action plays in non--linear $\sigma$--model theory. This soon called attention to the fact that in QFT the Ricci flow is naturally embedded into a more general geometric flow, the renormalization group flow for non--linear $\sigma$ models, which, even if mathematically ill-defined, provides an interpretation for the Ricci flow which is open to generalizations. In particular, the physics of non--linear $\sigma$ models shows that near a curvature singularity the renormalization group flow is no longer approximated by the Ricci flow but it should be replaced  by a geometric flow comprising new fields, coupled to geometry, that may give rise to a non--singular continuation of the flow. There are strong reasons to believe that the relevant singularity--suppressing contribution to this new dynamics does not come from the higher order loop expansions of the renormalization group, (providing corrections to the Ricci flow which are proportional to powers of the Riemann tensor and of its derivatives, and which may even enhance the formation of singularities  by introducing more reaction terms into the game). Rather, it is the physics of strong coupling that seems to play a role in singularity avoidance \cite{Douglas} by providing mass--generating mechanisms that transmute a collapsing solution of the Ricci flow (\emph{e.g.} a shrinking soliton) into a massive theory. The existence of a mass gap will imply that the shrinking soliton disappears. Alternatively, we may have a supersymmetry--driven mechanism generating a topological transmutation allowing the flow to be continued through the singular geometry. Renormalization group is too much a circle of idea rather than a theory, and thus it is not yet clear to what extent these indications may actually help in simplifying the actual, \emph{Ricci flow$+$surgery}, proof of the Geometrization theorem.  We are thus forced    
to dismantle the narrative, in favor of a  mathematical truth unshaped by physics, and conclude these introductory remarks by stressing that it is in QFT that the Ricci flow is paying back. Indeed, the understanding and the classification of singularities and the role of Ricci solitons, has provided  new strategies in the analysis of the landscape  of the  spaces of quantum field theories. We offer our own strategy which emphasizes the important role played by Ricci flow singularities and provide what we consider  a natural mechanism for extending, beyond the weak coupling regime, the embedding of the Ricci flow into the renormalization group flow. In any case, we leave the reader the choice of which track to follow in studying the connection between  renormalization of non--linear $\sigma$ models and Ricci flow theory. The amount of (intentional) and fortuitous inconsistencies in our presentation stresses the urge, for those aiming to a deep understanding of every small detail of this connection, to read  the original work by 
Daniel Friedan \cite{DanThesis,DanPRL, DanAnnPhys}, Sasha Zamolodchikov \cite{Zamo}, and Arkady Tseytlin \cite{Tseytlin1, Tseytlin2}. Prescient remarks \cite{Onofri}, and recent results are discussed in many papers, a fine selection being  \cite{Bakas, Bakas2, Bakas3}, \cite{Gegenberg}, \cite{Guenther}, \cite{Oliynyk}, \cite{Tseytlin}. 
%\begin{figure}[h]
%\begin{center}
%\includegraphics[scale=.7]{fig1}
%\caption{The map $\phi$ and the coupling  field $f$.}
%\end{center}
%\end{figure}

\section{QFT and the Renormalization Group flow}

To place the arguments to follow in a natural geometrical context, let us
denote by $\Sigma$ and $M$ smooth Riemannian manifolds with $\dim\Sigma=2$, $\dim M=n\geq2$, and let $Map\;(\Sigma,M):=\left\{\phi: \Sigma\rightarrow  M\right\}$ be the associated space of continuous maps. For simplicity of exposition we assume here that both the surface $\Sigma $ and the manifold $M$ are compact and oriented, and to allay anxiety we may suppose that $\phi \in Map\;(\Sigma,M)$ is of Sobolev class $W^{1,2}(\Sigma,M )$. In such a framework, let us consider the set of  natural Lagrangians on $Map(\Sigma,M)$ defined by some finite order jet of mappings $\phi\rightarrow \mathcal{L}(\phi,\alpha)$ from  $Map(\Sigma,M)$ to the space of (smooth) functions $C^{\infty }(\Sigma,{R})$, which are invariant under the diffeomorphism groups $\mathcal{D}iff(\Sigma )$, and $\mathcal{D}iff(M)$, and depend on a set of geometrical fields defined on $M$, collectively denoted by $\alpha$, and  which play the role of couplings parameters of the theory, 
(Fig. 1 and Fig. 2). Note that the
set of such coupling fields, $\mathcal{C}$, is itself an infinite dimensional space of geometrical origin.\\
The action associated with any such a Lagrangian is defined by $\mathcal{S}[\phi;\alpha]=\int_{\Sigma }\mathcal{L}(\phi,{\alpha})\,d\mu _{\Sigma }$, where $d\mu _{\Sigma }$ is a measure on $\Sigma $.  For instance, if $(\Sigma, \gamma)$ is a Riemannian $2$-dimensional surface with metric $\gamma$ and $(M,g)$ is a $n$--dimensional Riemannian manifold with metric $g$, then a typical natural lagrangian we may wish to consider is
$$ 
\mathcal{L}(\phi,\alpha)= a^{-1}\,\left[ tr_{\gamma(x)}\,(\phi^{*}\,g)+\,a\,f(\phi)\,\mathcal{K}\right] \;,
$$
where $tr_{\gamma(x)}\,(\phi^{*}\,g):=\gamma^{\mu\nu}(x)\,\partial _{\mu}\phi^{i}(x)\partial_{\nu}\phi^{j}(x)\,g_{ij}(\phi(x))$, $x\in \Sigma $, $\mu,\nu=1,2$, $i,j=1,\ldots,n$. Here  $a>0$ is a parameter with the dimension of a length squared, $f:M\to R$ is a smooth functions on $M$, and $\mathcal{K}$ is the Gaussian curvature of $(\Sigma,\gamma)$. 
%\begin{figure}[h]
%\begin{center}
%\includegraphics[scale=.7]{fig2}
%\caption{Spaces of maps $Map(\Sigma,M)$ parametrized by the space of couplings $\mathcal{C}$.}
%\end{center}
%\end{figure}
Since the dynamical field on $\Sigma $ is $\phi\in Map(\Sigma,M)$, the remaining field $f\in C^{\infty }(M,R)$, together with the metric tensor $g$ of $M$, play here the role of  point dependent coupling parameters $\alpha$ on $M$, \emph{i.e.}, 
$$
\alpha=a^{-1}\,\left(g,\, a\,f\right)\;,
$$
controlling the energetic of the action 
$$
\mathcal{S}[\phi;\alpha]=a^{-1}\,\int_{\Sigma }\left[ tr_{\gamma(x)}\,(\phi^{*}\,g)+\,a\,f(\phi)\,\mathcal{K}\right]\,d\mu _{\gamma }\;,
$$
where $d\mu _{\gamma }$ is the Riemannian volume element on $(\Sigma ,\gamma)$. 
Explicitly, $a^{-1}\,g$ sets the scale of $\phi(\Sigma )$ as seen in $(M,g)$, whereas the scalar field $f\in C^{\infty }(M,R)$, the \emph{dilaton}, provides the intrinsic scale of $(\Sigma ,\gamma)$, (a parameter, this latter, not captured by the energy density  $a^{-1}\,tr_{\gamma(x)}\,(\phi^{*}\,g)$ of the map $\phi\in Map(\Sigma,M)$ since this term is, for $dim \Sigma=2$, conformal invariant). Further position dependent coupling terms could have been added to the above action, (in particular
$a^{-1}\, \int_\Sigma U(\phi)\;d\mu_\gamma$ and $ a^{-1}\,\int_\Sigma \phi^*\omega $
 where $U\in C^{\infty }(M,R)$ and $\omega\in C^{\infty }(M,\wedge  ^{2}\,TM^{*})$ is a 2-form on $M$).\\
 
 \subsection{The space of natural Lagrangians}
 In classical physics (and in geometry) we typically fix our attention on a  fixed subset $\alpha_0\in \mathcal{C}$ of the possible coupling fields. For instance, in the above example we may set $\alpha_{0}=a^{-1}\,\left(g_{0},\, 0\right)$, and look for maps $\phi\in Map(\Sigma,M)$  minimizing $\mathcal{S}[\phi;\alpha_{0}]$, (\emph{viz.}, harmonic maps in $(M, g_{0})$ ). It is also often useful to consider  the same variational problem when we change the coupling fields $\alpha$ in a neighborhhod of the given $\alpha_{0}$, \emph{e.g.},  $\alpha_{0}\mapsto\alpha_{0}+\delta \alpha=a^{-1}\,\left(g_{0}+h,\,0+ a\,f\right)$, where the symmetric bilinear form $h\in C^{\infty }(M,\otimes ^{2}TM^{*})$ and the scalar field $f\in C^{\infty }(M,R)$ are small in a suitable norm. This allows us to study fluctuations and stability issues around the given extremizing fields $\phi\in Map(\Sigma,M)$ when we (adiabatically) change the geometry of the target manifold $(M,g)$.   This is a natural procedure which
characterizes the action $\mathcal{S}[\phi;\alpha]$ as a deformation of the fiducial $\mathcal{S}[\phi;\alpha_{0}]$. We write this deformation in general form as
\begin{equation}
\mathcal{S}[\phi;\alpha]=\mathcal{S}[\phi;\alpha_{0}] + \sum_{i\geq 1}\int_{\Sigma}\,\mathcal{O}_{i}(\phi,\alpha_{i})\;,
\label{SO}
\end{equation}
where we have denoted by $\mathcal{O}_{i}(\phi,\alpha_{i})$ the distinct terms in the lagrangian density $\mathcal{L}(\phi,\alpha)$ defining the deformations associated to the perturbed coupling fields $\alpha_{i}$. For instance, in the example which is presciently accompanying us, we may consider the deformation of the harmonic map action $\mathcal{S}[\phi;\alpha_{0}]= a^{-1}\,\int_{\Sigma }\left[ tr_{\gamma(x)}\,(\phi^{*}\,g_{0})\right]\,d\mu _{\gamma }$ defined by
$$
\mathcal{S}[\phi;\alpha]=\mathcal{S}[\phi;\alpha_{0}]+a^{-1}\,\int_{\Sigma }\left[ tr_{\gamma(x)}\,(\phi^{*}\,h)\right]\,d\mu _{\gamma }
$$
$$
+a^{-1}\,\int_{\Sigma }\,a\,f(\phi)\,\mathcal{K}\,d\mu _{\gamma }+a^{-1}\, \int_\Sigma U(\phi)\;d\mu_\gamma+a^{-1}\,\int_\Sigma \phi^*\omega\;,
$$
where $h\in C^{\infty }(M,\otimes ^{2}TM^{*})$, $f\in C^{\infty }(M,R)$, $U\in C^{\infty }(M,R)$, and $\omega \in C^{\infty }(M,\wedge  ^{2}TM^{*})$ are the (perturbing) fields. Note that $\mathcal{S}[\phi;\alpha_{0}]$ is invariant under the conformal transformation $(\Sigma ,\gamma_{\mu\nu})$ $\mapsto (\Sigma ,e^{-\psi}\,\gamma_{\mu\nu})$, $\psi\in C^{\infty }(\Sigma ,R)$. This symmetry is preserved by the perturbing fields $h$ and $\omega $, but is broken by the fields $f$ and $U$.  Further coupling fields can be introduced as long as the target Riemannian manifold $(M,g)$ is endowed with special geometrical structures (\emph{e.g.}, associated with  the presence of supersymmetries). At this point, it is also important to stress that classically the type of coupling fields $\delta\alpha$ which are compatible with the given $\mathcal{S}[\phi;\alpha_{0}]$ is dictated by the symmetry assumptions on $\mathcal{S}[\phi;\alpha_{0}]$ we wish to be preserved or broken by the perturbations. This is no longer trues in QFT where symmetries may be dynamically broken or generated by the spectrum of quantum fluctuations.  In any case,  it is somewhat natural to interpret the coupling fields so introduced 
$\alpha_{1}:=a^{-1}\,h$, $\alpha_{2}:=a^{-1}\,(af)$,  $\alpha_{3}:=a^{-1}\,U$, and $\alpha_{4}:=a^{-1}\,\omega $ as a sort of \emph{coordinates} for $\mathcal{S}[\phi;\alpha]$ in a neighborhood of the fiducial  $\mathcal{S}[\phi;\alpha_{0}]$, (Fig. 3). In a highly formal way we may think that this coordinatization provides a differentiable structure of a sort on the formal space $\mathcal{ACT}\,\,(\Sigma,M;\mathcal{C})$ of actions associated with natural Lagrangians on $Map(\Sigma,M)\times\mathcal{C}$, \emph{i.e.},
\begin{eqnarray}
&&\left.\mathcal{ACT}\,\,(\Sigma,M;\mathcal{C})\right|_{S_{0}-Patch}\,":="\\
\nonumber\\
&&\left\{(\ldots,\mathcal{O}_{i}(\phi,\alpha_{i})\ldots)\,|\;\mathcal{S}[\phi;\alpha]=\mathcal{S}[\phi;\alpha_{0}] + \sum_{i\geq 1}\int_{\Sigma}\,\mathcal{O}_{i}(\phi,\alpha_{i}) \right\} \notag\;.
\end{eqnarray}
This space, as formal as it may appear, relates naturally to Euclidean Quantum Field Theory where one is forced, by the very nature of the quantization procedure, to introduce a running energy scale parameter $t$ and a collection of special tangent vectors  $\frac{\partial }{\partial t}\,\alpha_{i}$, describing the variations of the coupling fields with the energy scale $t$, which formally provide a distinguished semi--flow in $\mathcal{C}$.
%\begin{figure}[h]
%\begin{center}
%\includegraphics[scale=.7]{fig3}
%\caption{A \emph{coordinatization} of the deformations of a given fiducial action.}
%\end{center}
%\end{figure}
\subsection{An informal geometrical view to renormalization}
In the above variational set up, the fluctuations in the field $\phi \in Map(\Sigma,M)$ govern at most a second order neighborhood of the configuration $\phi_0$ extremizing $\mathcal{S}[\phi;\alpha_{0}]$. In quantum physics the situation changes drastically and, at least in principle, all possible fluctuations in $\phi$ should matter. In such a setting,  the role of the action $\mathcal{S}[\phi;\alpha]$ is to provide a bias with which we weight the various configurations $\phi\in \mathcal{S}[\phi;\alpha_{0}]$ accessible to the system. The biasing mechanism is through quantum interference (QFT proper) or through stochastic  analysis (Euclidean QFT). In a (very few) favourable cases, the two mechanisms are related (axiomatic QFT) and the resulting theory allows for a formalism which is rather appealing in its mathematical structure. A good example which is worthwhile to keep in mind, and which to some extent fits nicely in our geometrical set up is the case when $\Sigma$ is the circle and $(M, g,\,x_0)$ is a pointed Riemannian manifold, \emph{i.e.}, the loop space $Map(S^1,M)$. This space is naturally endowed with a probability measure, the pinned Wiener  measure $\mathcal{W}_{x_0}(M)$ on continuous paths in $M$ starting and ending at some fixed point $x_0\in M$, and it is the framework appropriate for blending quantum mechanics with the Riemannian geometry of  $(M,g)$.  If the target manifold $M$ is a compact Lie group, (endowed with a bi--invariant Riemannian metric), then the measure space $\left\{Map(S^1,M),\,\mathcal{W}_{x_0}(M) \right\}$ is invariant under the flow induced by $W^{1,2}$ vector fields on $Map(S^1,M)$ \cite{Driver}. This is basically an extension of the Cameron--Martin theorem  which characterizes the path space $\mathcal{P}:=\{\eta\in C\left([0,1],\mathbb{R}^{n}\right)\,|\, \eta(0)=0\}$, endowed with the standard Wiener measure 
$\mathcal{W}_{0}$, (Fig. 4) and according to which the mapping $\mathcal{P}\rightarrow \mathcal{P}$, defined  by $\eta\mapsto \eta+f$ with $f\in \mathcal{P}$, $f(0)=0$, preserves (up to a density) the measure space $\{\mathcal{P},\, \mathcal{W}_{0}\}$ iff $\int_0^1\,|\frac{d}{ds}\,f|^2\,ds<\infty $. Explicitly, one defines $\mathcal{H}:=\{f\in \mathcal{P}\,|\, \frac{d}{ds}\,f \;exists\;a.e.\; and\; satisfies\; \int_0^1\,|\frac{d}{ds}\,f|^2\,ds<\infty \}$. This Hilbert space is densely embedded in $\mathcal{P}$, (however $\mathcal{W}_{0}(\mathcal{H})=0$), and can be identified with the tangent space $T_{\eta}\mathcal{P}$ to $\{\mathcal{P},\, \mathcal{W}_{0}\}$.  \\
%\begin{figure}[h]
%\begin{center}
%\includegraphics[scale=.7]{fig4}
%\caption{Wiener measure on path space over a Riemannian manifold}
%\end{center}
%\end{figure}
In our case, the Euclidean QFT of  relevance  is  characterized by the set of correlations, among the values $\{\phi(x_{i})\}\in (M)^{k}$ that the fields may attain at distinct marked points $x_{1},\ldots,x_{k}\,\in\Sigma$, formally defined by 
\begin{equation}
Z\left[\phi(x_{i});\alpha \right]"\doteq"\frac{1}{Z_{0}} \int_{\{Map(\Sigma,M)\}}D_{\alpha}[\phi]\,(\phi(x_{1})\ldots \phi(x_{i})\ldots)\;  e^{-S[\phi; \alpha] }\;,
\label{correlations}
\end{equation}
where $D_{\alpha}[\phi] $ is a  functional measure on $Map(\Sigma,M)$, possibly depending on the couplings $\alpha\in \mathcal{C}$, and $Z_{0}$ is a \emph{normalization constant} typically chosen so that 
${Z_{0}}^{-1}\,D_{\alpha}[\phi]$ is a probability measure, (Fig. 5). In particular, if, according to (\ref{SO}), we consider $S[\phi; \alpha]$ as a deformation of a fiducial $S[\phi; \alpha_{0}]$,\;\emph{i.e.}, $\mathcal{S}[\phi;\alpha]=\mathcal{S}[\phi;\alpha_{0}] + \sum_{a\geq 1}\int_{\Sigma}\,\mathcal{O}_{a}(\phi,\alpha_{a})$, then an expression of the structure (\ref{correlations}) follows from observing that we can write 
$$
\int_{\{Map(\Sigma,M)\}}D_{\alpha}[\phi]\;  e^{-S[\phi; \alpha] }= \int_{\{Map(\Sigma,M)\}}D_{\alpha}[\phi]e^{-S[\phi; \alpha_0] }\,\prod _{a} \int_{\Sigma}\mathcal{O}_{a}(\phi,\alpha_{a})\;,  
$$
where $\mathcal{O}_{a}(\phi,\alpha_{a})$ are to be promoted to (operator--valued) local distributions (supported on distinct points $x_a\in \Sigma $), and the $\prod_{a}$ is suitably ordered. Thus, in such a setting one typically assumes ${Z_{0}}:=\,\int_{\{Map(\Sigma,M)\}}D_{\alpha}[\phi]e^{-S[\phi; \alpha_0] }$. Notice also that the probability measure so defined formally induces on the coupling space $\mathcal{C}$ a covariance
\begin{equation}
\mathcal{G}(\alpha_i,\alpha_j):=\frac{1}{Z_{0}}\, \int_{\{Map(\Sigma,M)\}}D_{\alpha}[\phi]e^{-S[\phi; \alpha_0] }\,\int_{\Sigma}\mathcal{O}_{i}(\phi,\alpha_{i})\, \int_{\Sigma}\mathcal{O}_{j}(\phi,\alpha_{j})\;,  
\label{Zamometric} 
\end{equation}
which, if positive, turns $\mathcal{C}$ into a measure space, $\left\{\mathcal{C};\,D[\mathcal{G}]\right\}$, (this covariance is often called the Zamolodchikov metric in 2D QFT).\\
%\begin{figure}[h]
%\begin{center}
%\includegraphics[scale=.9]{fig5}
%\caption{Correlations over fluctuating surfaces}
%\end{center}
%\end{figure}
Although rigorous bona--fide functional measure on the map spaces $Map(\Sigma,M)$ can be  introduced (cf.\, \cite{Taubes, Weitsman, Leandre}),  
the somewhat fanciful expressions defined above hardly makes  sense by themselves, even at a physical level of rigor,  if we do not devise a way of controlling the spectrum of fluctuations of the fields $\phi\in Map(\Sigma,M)$. In particular, it is not obvious how to introduce a subspace of $\{Map(\Sigma,M)\, e^{-S[\phi; \alpha] }\,D_{\alpha}[\phi] \}$ playing the role that the Cameron--Martin tangent space $\mathcal{H}$ has in the case of the Wiener measure space $\{\mathcal{P},\, \mathcal{W}_{0}\}$. Indeed, one fundamental problem concerning (\ref{correlations}) is to introduce a filtration in $\{Map(\Sigma,M)\,, e^{-S[\phi; \alpha] }\,D_{\alpha}[\phi] \}$ , parametrized by a length scale $t$, (the only scale of measurement significant in a relativistic quantum theory). This filtration allows to control the way (\ref{correlations})  behaves under scale--dependent transformations of the fields $\phi\in Map(\Sigma,M)$ and of the couplings $\alpha\in \mathcal{C}$, (Fig. 6). 
Thus, a basic ingredient in any such a QFT is the search for a flow, (\emph{renormalization group flow}),
\begin{eqnarray}
\mathcal{RG}_{t}:\,[Map(\Sigma,M)\times\mathcal{C}]&\longrightarrow& [Map(\Sigma,M)\times\mathcal{C}]\label{reflow}\\
(\phi,\alpha)\;\;\;\;\;\;\;&\longmapsto& \mathcal{RG}_{t}(\phi,\alpha)=({\phi}_{t};{\alpha}(t))\;,\nonumber
\end{eqnarray}
which, as we vary the scale of distances $t$ at which we probe the Riemannian surface $\Sigma$, allows to tame the energetics of the fluctuations of the fields $\phi : \Sigma\to M$ in terms of the couplings $\alpha\mapsto\alpha(t)$. In order to describe this procedure in physical terms, select two scales of distances, say $\Lambda^{-1}$ and $\Lambda'^{-1}$, (one can equivalently interpret $\Lambda$ and $\Lambda'$ as the respective scales of momentum in the spectra of field fluctuations), with   $\Lambda'^{-1}>\Lambda^{-1}$.  The general idea, central in K.G. Wilson's analysis of the the renormalization group flow, is to assume that if the action $S[\phi_{\Lambda }; \alpha(\Lambda )]\in \mathcal{ACT}\,\,(\Sigma,M;\,\mathcal{C})$ describes the theory at  a cut--off scale $\Lambda^{-1}$, then  there is a map 
%\begin{figure}[h]
%\begin{center}
%\includegraphics[scale=.9]{fig6}
%\caption{Averaging fluctuations over length scales $<t$.}
%\end{center}
%\end{figure}
\begin{eqnarray}
{\widetilde{\mathcal{RG}}}_{\Lambda\Lambda'}:\mathcal{ACT}\,\,(\Sigma,M;\,\mathcal{C})&\longrightarrow &\;\;\;\;\;\;\mathcal{ACT}\,\,(\Sigma,M;\,\mathcal{C})\;,\\
S[\phi_{\Lambda }; \alpha(\Lambda )]\;\;\;\;&\longmapsto & S[\phi_{\Lambda' }; \alpha(\Lambda ')]\doteq    {\widetilde{\mathcal{RG}}}_{\Lambda\Lambda'}\,S[\phi_{\Lambda }; \alpha(\Lambda )]\notag 
\end{eqnarray}
such that (Fig. 7) the action $S[\phi_{\Lambda' }; \alpha(\Lambda ')]\doteq    {\widetilde{\mathcal{RG}}}_{\Lambda\Lambda'}\,S[\phi_{\Lambda }; \alpha(\Lambda )]$     provides the effective theory at scale $\Lambda'^{-1}$, obtained upon suitably averaging field--fluctuations in moving from the distance scale $\Lambda^{-1}$ to the scale $\Lambda'^{-1}$, (note that with respect to \cite{romp} we are eliminating a few idiosyncratic $*$ from the notation). Such a map is required to satisfy the semigroup property   ${\widetilde{\mathcal{RG}}}_{\Lambda\Lambda''}={\widetilde{\mathcal{RG}}}_{\Lambda'\Lambda''}\circ {\widetilde{\mathcal{RG}}}_{\Lambda\Lambda'}$ for all $\Lambda''<\Lambda'$. This formal (semi)-flow, if exists, induces a corresponding flow  on $Map(\Sigma,M)\times\mathcal{C}$ 
\begin{eqnarray}
\mathcal{RG}_{\Lambda\Lambda'}:\,[Map(\Sigma,M)\times\mathcal{C}]&\longrightarrow& [Map(\Sigma,M)\times\mathcal{C}]\label{reflow2}\\
(\phi_{\Lambda },\alpha(\Lambda ))\;\;\;\;\;\;\;&\longmapsto& \mathcal{RG}_{\Lambda\Lambda'}(\phi_{\Lambda },\alpha(\Lambda ))=({\phi}_{\Lambda'};{\alpha}(\Lambda'))\;,\nonumber
\end{eqnarray}
by requiring the natural commutativity relation 
$$
{\widetilde{\mathcal{RG}}}_{\Lambda\Lambda'}\,S[\phi_{\Lambda }; \alpha(\Lambda )]=S[\mathcal{RG}_{\Lambda\Lambda'}(\phi_{\Lambda },\alpha(\Lambda ))]\;,
$$
holds.
%\begin{figure}[h]
%\begin{center}
%\includegraphics[scale=.7]{fig7}
%\caption{Renormalization as a map in the space of actions.}
%\end{center}
%\end{figure}
 Note that formally, under the action of  ${\mathcal{RG}}_{\Lambda\Lambda'}$ we can either pull--back or push--forward the measure $e^{-S[\phi; \alpha] }\,D_{\alpha}[\phi]$. Given the physical meaning of the renormalization group flow, the push--forward should be perhaps more appropriate in a measure--theoretic sense, however, for simplicity we shall use the pull--back measure ${\mathcal{RG}}_{\Lambda\Lambda'}^{*}\,(D_{\alpha}[\phi])\;  e^{-\widetilde{\mathcal{RG}}_{\Lambda\Lambda'}\,S[\phi; \alpha]}$. Indeed, in order to characterize the flow $\mathcal{RG}_{\Lambda\Lambda'}$ we \emph{assume} that it should leave the measure space $\{Map(\Sigma,M)\times \mathcal{C},\, e^{-S[\phi; \alpha] }\,D_{\alpha}[\phi] \}$ quasi--invariant in a suitable sense. 
The idea is roughly the following: suppose that, at least for $({\Lambda'}\setminus {\Lambda})$ small enough, we can put the pulled--back functional measure ${\mathcal{RG}}_{\Lambda\Lambda'}^{*}\,(D_{\alpha}[\phi])\;  e^{-\widetilde{\mathcal{RG}}_{\Lambda\Lambda'}\,S[\phi; \alpha]}$ in the same form as the original functional measure $D_{\alpha}[\phi]\;  e^{-S[\phi; \alpha] }$, except for a small modification of the couplings $\alpha$. Explicitly, let $\Lambda'=e^{-\epsilon}\,\Lambda$, with $0<\epsilon<1$, and assume that for every such $\epsilon$ there exists a corresponding coupling $\alpha+\delta\,\alpha$ such that the following identity holds
\begin{equation}
\mathcal{RG}_{\epsilon}^{*}\,(D_{\alpha}[\phi])\;  e^{-\widetilde{\mathcal{RG}}_{\epsilon }\,S[\phi; \alpha]}= D_{\alpha+\delta\alpha}[\phi]\;  e^{-S[\phi; \alpha+\delta\alpha] }\;,
\label{infrenor}
\end{equation}
where we have denoted $\widetilde{\mathcal{RG}}_{\epsilon}$ the action of the map $\widetilde{\mathcal{RG}}_{\Lambda\Lambda'}$ for $\Lambda'=e^{-\epsilon}\,\Lambda$.
In other words, we assume that an infinitesimal change in the cutoff can be completely \emph{absorbed} in an infinitesimal change of the couplings. If this equation  is valid at least to some order in $\epsilon$,  we can iteratively use it to see how $\alpha$ is affected by a finite change of the cutoff. If we set $t\doteq\,-\ln\,({\Lambda'}\setminus {\Lambda})$, then the map $\mathcal{RG}_{t}$ so induced by $\widetilde{\mathcal{RG}}_{\Lambda\Lambda'}$ on $[Map(\Sigma,M)\times\mathcal{C}]$, as $t$ varies, is the renormalization group flow $\mathcal{RG}_{t}$ introduced above, (see (\ref{reflow})). Since  $[Map(\Sigma,M)\times\mathcal{C}]$ is non--linear, the infinitesimal quasi--invariance described by (\ref{infrenor}) is what we may reasonably expect to replace the quasi--invariance characterizing linear measure spaces such as $\{\mathcal{P},\, \mathcal{W}_{0}\}$. This infinitesimal quasi--invariance yields for what is basically an integration by parts formula characterizing a set of distinguished tangent vector fields to the measure space $\left\{\mathcal{C};\,D[\mathcal{G}]\right\}$. 
To show how this comes about, let us consider a scale interval $-\epsilon \leq t \leq \epsilon $, for $\epsilon>0$, and  assume  that  the associated functional measure $D_{\alpha}[\phi]\;  e^{-S[\phi; \alpha] }$ has natural transformation properties under $\mathcal{RG}_{t}$, \emph{i.e.}, 
\begin{align}
&&\int_{\mathcal{RG}_{t}\{Map(\Sigma,M)\}}D_{\alpha(t)}[\phi_t]\;  e^{-S[\phi_t; \alpha(t)] } \label{renorm}\\
\notag\\
&&= \int_{\{Map(\Sigma,M)\}}\,\mathcal{RG}_{t}^{*}\,(D_{\alpha}[\phi])\;  e^{-\widetilde{\mathcal{RG}}_{t}\,S[\phi; \alpha] }\notag\;.
\end{align}
The strategy is to exploit (\ref{infrenor}) by evaluating, along the $\mathcal{RG}_{t}$ map, the flow derivative $\frac{d}{d t}\,Z[\alpha(t)]$ at the generic scale $t$, where 
$$
Z[\alpha(t)]\doteq \int_{\mathcal{RG}_{t}\{Map(\Sigma,M)\}}D_{\alpha(t)}[\phi_t]\;  e^{-S[\phi_t; \alpha(t)] }\;. 
$$
Denoting, from notational ease, $(\Sigma,M)_{t}:=\mathcal{RG}_{t}\{Map(\Sigma,M)\}$,  we compute at a very (in)formal level
\begin{align}
\label{zdiff}
&&\frac{d}{d t}\,Z[\alpha(t)]=\lim_{\epsilon \rightarrow 0}\,\frac{1}{\epsilon }\left[\int_{(\Sigma,M)_{t+\epsilon }}D_{\alpha(t)}[\phi_t]\;  e^{-S[\phi_t; \alpha(t)] }\right.\\
&&\left.-\,\int_{(\Sigma,M)_{t}}D_{\alpha(t)}[\phi_t]\;  e^{-S[\phi_t; \alpha(t)] }\right]\nonumber\\
\nonumber\\
&&=\lim_{\epsilon \rightarrow 0}\,\frac{1}{\epsilon }\left[\int_{\mathcal{RG}_{\epsilon }((\Sigma,M)_{t})}D_{\alpha(t)}[\phi_t]\;  e^{-S[\phi_t; \alpha(t)] }\right.\nonumber\\
&&\left.-\,\int_{(\Sigma,M)_{t}}D_{\alpha(t)}[\phi_t]\;  e^{-S[\phi_t; \alpha(t)] }\right]\nonumber\\
\nonumber\\
&&=\int_{(\Sigma,M)_{t}}\lim_{\epsilon \rightarrow 0}\,\frac{1}{\epsilon }\left[\mathcal{RG}^*_{\epsilon }[D_{\alpha(t)}[\phi_t]]\;  
e^{-\widetilde{\mathcal{RG}}_{\epsilon }\,S[\phi_t; \alpha(t)]}\right.\nonumber\\
&&\left.-D_{\alpha(t)}[\phi_t]\;  e^{-S[\phi_t; \alpha(t)] }     \right]
\nonumber\\
&&=-\,\int_{(\Sigma,M)_{t}}\,\beta(\alpha(t))\frac{\partial}{\partial \alpha(t)}\,
\left(D_{\alpha(t)}[\phi_t]\;  e^{-S[\phi_t; \alpha(t)] }\right)\nonumber\;,
\end{align}
where we have introduced the $\beta$--flow \emph{vector field} (Fig. 8) on the space of couplings $\mathcal{C}$
\begin{equation}
\beta(\alpha(t)) \doteq -\frac{\partial}{\partial t}\,\alpha(t)\;,
\label{betaflow}
\end{equation}
 and where we have exploited the semigroup property of the flow and the scaling hypothesis (\ref{infrenor}).
Since the integration is over $\mathcal{RG}_{t}\{Map(\Sigma,M)\}$, we can formally extract the operator $\beta(\alpha(t))\frac{\partial}{\partial \alpha (t)}$   from under the functional integral, and 
 rewrite the relation  (\ref{zdiff}) as 
\begin{equation}
\left\{\frac{d}{d t} +\beta(\alpha(t))\frac{\partial}{\partial \alpha(t)}\right\}\,
Z_t[\alpha]=0\;.
\label{renormflow}
\end{equation}
%\begin{figure}[h]
%\begin{center}
%\includegraphics[scale=.7]{fig8}
%\caption{The geometry of the beta function.}
%\end{center}
%\end{figure}
Roughly speaking (\ref{renormflow}) says is that if we rescale  distances in $\Sigma$ by a factor $e^{t}$ and at the same time we flow in the space of couplings along $\beta$ for a \emph{time} $t$, the theory we obtain looks the same as before. If the theory is, along the lines sketched above, renormalizable by a renormalization of the couplings, many of its properties can be desumed by the analysis of (\ref{betaflow}). 
In the framework so described, the renormalization group formally appears as a natural geometrical flow on the measure space  $\left\{Map(\Sigma,M)\times\mathcal{C};\,D_{\alpha(t)}[\phi_t]\right\}$ and the 
$\beta$--flow \emph{vector fields} (\ref{betaflow}) play a role analogous to the role played by the Cameron--Martin vector fields for Wiener measure space. In particular, since  $\left\{\mathcal{C};\,D[\mathcal{G}]\right\}$ is a measure space when endowed with the $D_{\alpha(t)}[\phi_t]$--induced covariance defined by the \emph{Zamolodchikov metric} $\mathcal{G}_{ij}$, (see (\ref{Zamometric})), it is natural to argue whether the  $\beta$-- vector fields are gradient of a suitable functional $\Phi (\alpha)$ with respect to $\mathcal{G}$: \emph{i.e.}, if $\beta(\alpha)=\mathcal{G}(\mathcal{D}\Phi (\alpha)\circ \nu)$, (here $\mathcal{D}\Phi (\alpha)\circ \nu$ denotes the linearization of $\Phi (\alpha)$ in the direction of the coupling perturbation $\nu$). This is a fascinating open issue, (see \emph{e.g.} \cite{Cappelli}),  deeply connected with a geometrical understanding of the  renormalization group in its role of averaging out fluctuations: being a gradient flow would avoid recurrent behaviors like limit cycles and strange attractors. It would also imply  that any such a functional  $\Phi (\alpha)$ is monotonically non--increasing along the flow and that the renormalization group flow is, as intuitively expected,  irreversible. A somewhat weaker form of such irreversibility is associated with the celebrated  Zamolodchikov's $c$--\emph{theorem} \cite{Zamo}, (actually more a conjecture than a theorem), which, under a unitarity condition, states that for 2D QFT there exists a function $c(\alpha)$ which is monotonically non--increasing along the renormalization group flow. The associated fixed points of the RG flow are conformal field theories (CFT), corresponding to which $c(\alpha)$ reduces to the corresponding central charge $c$. \\
The connections between the Renormalization Group flow, stochastic analysis, gradient flows,  $c$--theorems and Ricci flow become rather manifest when one computes explicitly the beta flow for the QFT associated with the harmonic map action we have been using as our guiding example.

\section{Ricci flow and the renormalization group}

Let us consider again the  action 
\begin{equation}
\mathcal{S}[\phi;\alpha] \doteq a^{-1}\,\int_{\Sigma }\left[ tr_{\gamma(x)}\,(\phi^{*}\,g)\right]\,d\mu _{\gamma }
=a^{-1}\,\int_\Sigma\gamma^{\mu\nu}\partial_\mu\phi^i\partial_\nu\phi^j\,g_{ij}\,d\mu_\gamma\;,
\label{sigmamod}
\end{equation}
the  critical points of which are harmonic maps of the Riemann surface $(\Sigma,\gamma) $ into $(M,g)$. We  assume explicitly that $\Sigma$ is the flat torus ${T}^2={R}^2/{Z}^2$,  $\gamma_{\mu\nu}=\delta_{\mu\nu}$, and emphasize once more the fact that at each point $x\in\Sigma$ the metric $a^{-1}\,g(\phi(x))$ plays the role of the coupling constants for the fields $\phi(x)$ of the theory.   In quantum theory, this fiducial action together with its possible deformations, describe a family of $2$--dimensional QFTs known as non--linear $\sigma$--models. They find applications ranging from  condensed matter physics to string theory, but of particular relevance for us is their role in providing a testing ground for exploring the landscape of the space of quantum field theories. For this reason, and also as a way of illustrating the role of the Ricci flow, we will discuss in some detail the renormalization group analysis of (\ref{sigmamod}). What follows is mostly taken from \cite{romp}, a mathematical digest of the  fine presentation of non--linear $\sigma $--model theory by K. Gawedzki \cite{Gaw}. 
Let me stress that we will limit our analysis to the quantum deformation of the theory involving only the metrical coupling $\alpha=a^{-1}\,g$. This is the situation generating the Ricci flow. The extension to the dilaton coupling field $f(\phi)$, (associated with the action term $a^{-1}\,\int_{\Sigma }\,a\,f(\phi)\,\mathcal{K}\,d\mu _{\gamma }$), relevant to Perelman's analysis, does not present a particular difficulty and we simply state the final results when needed.\\
%\begin{figure}[h]
%\begin{center}
%\includegraphics[scale=.7]{fig9}
%\caption{The parameter $a$ and the curvature of $|Riem(g)|$  set the scale at which $(\Sigma,\gamma)$ probes the ambient manifold $(M,g)$}
%\end{center}
%\end{figure}
\subsection{The QFT analysis of $S[\phi;\,\alpha]$}
Let us start by observing that, under the above assumptions, the space of couplings $\mathcal{C}$ can be identified with the infinite--dimensional stratified manifold  $\mathcal{C}=\frac{\mathcal{M}et(M)}{\mathcal{D}iff(M)\times R^{+}}$ of Riemannian structures on $M$, modulo overall length rescalings, where $\mathcal{M}et(M)$ denotes the cone of Riemannian metrics over $M$, $\mathcal{D}iff(M)$ is the group of diffeomorphisms of $M$, and where $R^{+}$ denotes the group of rescalings defined by $a\mapsto\lambda a$, for $\lambda$ a positive number. 
Since the true dimensionless coupling constant of the theory is
the ratio of the length scale of the target space 
metric $g_{ab}$ (\emph{i.e.}, its squared radius of curvature ${r^{2}_{curv}}$) to $a$, we may consider a point--like limit,
where the size of the surface $(\Sigma,\gamma)$ is much
smaller than the physical length scale of $(M,g_{ab})$, (Fig. 9). This implies that when curvature of target Riemannian manifold $(M,g)$ is small as seen by $\Sigma$, the measure  $D_{g}[\phi]\,e^{-S[\phi;\,\alpha]}$ is concentrated
around the minima of the fiducial action $S[\phi;\,\alpha]$, \emph{i.e.} the constant maps $x\to\phi(x)=\phi_{0}$, and we can
 control the nearly Gaussian fluctuations $\delta\phi$, (Fig. 10). With a slight abuse of language, it is typical to talk in this case of  perturbation
theory for small $a$, and say that the theory is perturbatively renormalizable in terms
 of the scale parameter $a$. \\
%\begin{figure}[h]
%\begin{center}
%\includegraphics[scale=.9]{fig10}
%\caption{The point--like limit: Nearly Gaussian fluctuations near constant maps}
%\end{center}
%\end{figure} 
A major technical difficulty in implementing a renormalization group procedure for the QFT associated with (\ref{sigmamod}) is that
the space of maps $Map(\Sigma, M)$ is a non--linear functional space. However, as we have seen above, in the weak--coupling limit, where the size of the surface $(\Sigma,\gamma)$ is much
smaller than the physical length scale of $(M,g_{ab})$ only fields fluctuating around a constant value $\phi_0\in M$ play a role. It follows that one can work in a geodesic ball $B(\phi_0,r)\subset M$, centered at the point $\phi_0$, with radius $$r<\min\left\{\frac{1}{3}\,inj\;(\phi_0),\,\frac{\pi }{6\,\sqrt{K}}  \right\}\;,$$
where $K$ is an upper bound to the sectional curvature of $(M,g)$, (we are adopting the standard convention of defining $\pi/2\,\sqrt{K}\doteq\infty$ when $K\leq0$), and $inj\;(\phi_0)$ denotes the injectivity radius of $(M,g)$ at $\phi_0$. Under these hypotheses, given $N$ independent copies $\{\phi_k:\Sigma\to B(\phi_0,r)\}_{k=1,\dots,N}$, of the field $\phi:\Sigma\to B(\phi_0,r)$,  one can define their \emph{center of mass}, (Fig. 11) 
\begin{equation}
\psi\doteq cm\, \left\{\phi_1,\ldots,\phi_N \right\}\;, 
\label{cm}
\end{equation}
as the minimizer of the function $F:M\to R$, defined by 
$$F(y)\doteq \frac{1}{2}\,\sum_{k=1}^{N}\,d_{g}^{2}(y,\phi_k)\;,$$
where $d_{g}^{2}(\circ ,\circ )$ denotes the distance function in $(M,g)$. Note that if $inj\;(y)>3r$ for all $y\in B(\phi_0,r)$, then the minimizer is unique and $cm\, \left\{\phi_1,\ldots,\phi_N \right\}\,\in B(\phi_0,2r)$\,\, \cite{Glickenstein}. 
%\begin{figure}[h]
%\begin{center}
%\includegraphics[scale=.7]{fig11}
%\caption{In the point--like limit we can define the center of mass $\psi$ of  $N$ indipendent  copies of the mapping $\phi$}
%\end{center}
%\end{figure}
 The idea is to describe the QFT, corresponding non--linear sigma model action (\ref{sigmamod}), by extracting the behavior of the (quantum) fluctuations of the maps $\phi:\Sigma\to B(\phi_0,r)$ around the \emph{background (or average) field} $\psi$ defined by the distribution of the center of mass  of a large ($N\to\infty$) number of independent copies of $\phi$ itself. Thus, we shall consider  $N$ fields $\{\phi_k:\Sigma\to B(\phi_0,r)\}$, indipendently distributed on $Map(\Sigma, B(\phi_0,r))\subset Map(\Sigma, B(\phi_0,2r))$ according to the measure  $D_{g}[\phi])\; e^{-(S[(\phi; \alpha)] }$.  The associated \emph{background field} $\psi: \Sigma\to B(\phi_0,2r)$, (\ref{cm}), then is distributed according to the law
\begin{equation}
P_N(\psi)= \int_{Map(\Sigma, B(\phi_0,2r))}\,\delta(cm\{\phi_j(x)\}-
\psi(x))\prod_{j=1}^N\,D_{g}[\phi_j]\; e^{-S[(\phi_j; \alpha)]}\;,
\label{distribution}
\end{equation}
where $\delta(cm\{\phi_j(x)\}-\psi(x))$ is the Dirac measure at  $\psi(x)\in Map(\Sigma, B(\phi_0,2r))$. If one considers the  exponential map, $\exp_{\psi}: T_{\psi}M\to B(\phi_0,2r)$, based at the center of mass $\psi\in B(\phi_0,2r)$, and writes $\phi_k=\exp_{\psi}\,(X_k)$,  $\{X_k\}_{k=1,\ldots,N}\in T_{\psi}M$, then (\ref{distribution}) can be reformulated in terms of fields taking values in the pull--back bundle  $\left.\psi^{*}\,TM\right|_{B(\phi_0,2r)}$. To this end, one introduces $N$ sections $\xi _k:\Sigma\to \psi^{*}\,TM$, with $\;\sum_{j=1}^N\xi_j=0$ (because $\exp_{\psi}\,(\cdot )$ is based at the center of mass $\psi$), and  such that that $X_k=\psi_{*}\,\xi _k$, 
(Fig. 12). Thus
\begin{equation}
\phi_k(x) = \exp_{\psi(x)}(\psi_{*}\,\xi_k(x))\;,
\end{equation}
and one can conveniently express (\ref{distribution}) as a functional integral, over the linear space of maps $Map\left(\Sigma, \psi^{*}\,TM\right)\doteq Map\left(\Sigma, \psi^{*}\,TM|_{B(\phi_0,2r)}\right)$, according to
\begin{equation}
\int_{Map\left(\Sigma, \psi^{*}\,TM\right)}\,\delta[\sum_{j=1}^N\xi_j]\prod_{k=1}^N\,e^{-S_{\psi}[\xi_k;\,\alpha]}
D^{\psi}_g\,[\xi_k]\;,
\end{equation}
where $S_{\psi}[\xi_k;\,\alpha]\doteq S[\exp_{\psi}(\psi_{*}\,\xi_k);\,\alpha]$ and $D^{\psi}_g\,[\xi_k]\doteq D_g\,[\exp_{\psi}(\psi_{*}\,\xi_k)]$.
By exploiting the formal Fourier representation of the functional Dirac--$\delta[\circ ]$ one can write
$$
\delta[\sum_{j=1}^N\xi_j]=\int_{Map^{*}(\Sigma, \psi^{*}\,TM)} \left[DJ\right]\,\exp\,i\, \langle J\cdot\sum_{j=1}^N\xi_j \rangle \;,
$$
where the pairing $\langle\circ ,\circ \rangle$ between $Map\left(\Sigma, \psi^{*}\,TM\right)$ and its dual $Map^{*}\left(\Sigma, \psi^{*}\,TM\right)$ is defined by the   $L^{2}$ inner product
$$
\langle J\cdot\sum_{j=1}^N\xi_j \rangle\doteq \int_{\psi^{*}\,TM|_{B(\phi_0,2r)}}\,(\psi^{*}g)_{\mu\nu}\,J^{\mu}\sum_{j=1}^N\xi^{\nu}_j\,\psi^{*}d\mu_{g}\;.
$$
Thus one can eventually express (\ref{distribution}) as
\begin{eqnarray}
\label{Jpartition}
&P_N(\psi)=\int\left[DJ\right]\,\int\,e^{i\,\langle J\cdot\sum_{j=1}^N\xi_j\rangle}\,\prod_{k=1}^N\,e^{-S_{\psi}[\xi_k;\,\alpha]}\,D^{\psi}_g\,[\xi_k]\\
\nonumber\\
&=\int\left[DJ\right]\,e^{N\,W_{\psi}(J)}\nonumber\;,
\end{eqnarray}
where we have introduced the characteristic functional of the functional measure $e^{-S_{\psi}[\eta ;\,\alpha]}\,D^{\psi}_g\,[\eta ]$,  $\eta\in\,Map\left(\Sigma, \psi^{*}\,TM\right)$, according to
\begin{equation}
e^{W_{\psi}(J)}\doteq\int_{Map(\Sigma, \psi^{*}\,TM)}\,e^{i\,\langle J\cdot\eta\rangle}\,e^{-S_{\psi}[\eta;\,\alpha]}\,D^{\psi}_g\,[\eta]\;.
\end{equation}
Note that one may provide an asymptotic expansion for $W_{\psi}(J)$ by Taylor expanding  $S_{\psi}[\eta;\,\alpha]$ around its minimum, (at $\eta=0$), and by separating the  Gaussian measure 
$D^{\psi}_g[\Xi]\doteq{e^{-\frac{1}{2}S_{\mu\nu}^{\psi}\eta^{\mu}\eta^{\nu}}D^{\psi}_g\,[\eta]}\backslash [{\int\,
e^{-\frac{1}{2}S_{\mu\nu}^{\psi}\eta^{\mu}\eta^{\nu}}D^{\psi}_g\,[\eta]}]^{-1}$, 
where $S_{\mu\nu\ldots}^{\psi}$ denotes the covariant derivatives of the action $S_{\psi}[\eta;\,\alpha]$ evaluated for $\eta=0$.
In this way we get
\begin{eqnarray}
&&e^{W_{\psi}(J)}=e^{-S_{\psi}[0;\,\alpha]}\,\left[\int\,e^{-\frac{1}{2}S_{\mu\nu}^{\psi}[0;\,\alpha]\eta^{\mu}\eta^{\nu}}D^{\psi}_g\,[\eta]\right]\;\times \\
\nonumber\\
&&\times \int\,D^{\psi}_g[\Xi]\,e^{i\,\langle J\cdot\eta\rangle}\,e^{-S_{\alpha\mu\nu}^{\psi}[0;\,\alpha]\eta^{\alpha}\eta^{\mu}\eta^{\nu}\,-\ldots\,}
\nonumber\;,
\end{eqnarray}
%\begin{figure}[h]
%\begin{center}
%\includegraphics[scale=.8]{fig12}
%\caption{The geometrical set up for discussing fluctuations around the background field $\psi$}
%\end{center}
%\end{figure}
 The expansion in power series of all the exponentials and the resulting term--by--term Gaussian integration provide the formal expression
\begin{align}
&&W_{\psi}(J)=-S_{\psi}[0;\,a^{-1}\,g]+\sum_{\Upsilon \in G}\,\frac{(a)^{l(\Upsilon )}}{|Aut(\Upsilon )|}
\,F_{\Upsilon }\left(S_{\psi},\,J\right)\label{Wfeynman}\\ 
&&+\,\ln\,\left(\int\,e^{-\frac{1}{2}S_{\mu\nu}^{\psi}[0;\,\alpha]\eta^{\mu}\eta^{\nu}}D^{\psi}_g\,[\eta] \right)\notag\;,
\end{align}
where $G$ denotes the set of isomorphism classes of connected graphs $\Upsilon $ without external lines and with $l(\Upsilon )$ loops. $|Aut(\Upsilon )|$ denotes the size of the corresponding automorphisms group. $F_{\Upsilon }\left(S_{\psi},\,J\right)$ is the Feynman amplitude  of each given $\Upsilon \in G$, computed by associating to $1$--leg vertices the current $J$, to each $n$--leg vertices, $n\geq3$, the interaction $S_{\alpha_1\ldots\alpha_n}^{\psi}[0;\,\alpha]$, and to any internal edge the propagator defined by $S_{\mu\nu}^{\psi}[0;\,\alpha]$. \\
\\
According to (\ref{Jpartition}) the large $N$ asymptotics of the distribution $P_N(\psi)$ of the background field $\psi$ is  provided by
$$
P_N(\psi)=e^{N\;\inf_{J}\;W_{\psi}(J)\,+\,o(N)}\;,
$$
where the $\inf$ is over all ${J\in\, Map^{*}(\Sigma, \psi^{*}\,TM)}$. As emphasized in\cite{Gaw}, 
$$\sup_{J}\;[\langle\zeta,J\rangle- W_{\psi}(J)]$$
 is the large deviation functional governing the $\mathcal{O}(N)$--fluctuations around $\zeta $, (in our case $\zeta =0$), in the distribution of $\{\xi_j\}_{j=1,\ldots,N}$, as compared to the $\mathcal{O}(\sqrt{N})$ Gaussian fluctuations sampled by the central limit theorem, (Fig. 13). Since $\sup_{J}\;[\langle\zeta,J\rangle- W_{\psi}(J)]$ is the Legendre transform of $W_{\psi}(J)$, it follows, from standard QFT, that $\sup_{J}\;[\langle\zeta,J\rangle- W_{\psi}(J)]$ can be identified with the \emph{effective action} associated with $W_{\psi}(J)$, \emph{i.e.} with the action functional whose corresponding partition function gives, at tree (classical) level, the full  characteristic functional $W_{\psi}(J)$. According to these remarks it follows that, for the non--linear sigma model (\ref{sigmamod}), the role of a \emph{background field effective action} is played by
$$\Gamma(\psi)\doteq \sup_{J}\;[-\,W_{\psi}(J)]\;.$$
Geometrically, this is the large deviation functional controlling the non--Gaussian fluctuations of the fields $\{\phi_j\}_{j=1,\ldots,N}$,  around a \emph{background (or classical) field} $\psi$ obtained as average of a large number of copies $\{\phi _{j}\}_{j=1,\ldots,N\to\infty}$ of the quantum field itself.\\
%\\
%\begin{figure}[h]
%\begin{center}
%\includegraphics[scale=.8]{fig13}
%\caption{Nearly Gaussian distribution of the fields $\eta$ in the linear space of maps  $Map(\Sigma,\psi^*TM)$. This distribution generates the %effective action for the background field $\psi$.}
%\end{center}
%\end{figure}
The full effective action $\Gamma(\psi)$ can be perturbatively defined, starting from the expansion (\ref{Wfeynman}) of $W_{\psi}(J)$,  by rewriting it in terms of 1--particle irreducible (1PI) graphs, (\emph{i.e.}, in terms of connected graphs  without bridges, where an edge $e$ of a connected graph $\Upsilon $ is said to be a bridge if the graph $\Upsilon \setminus e$ is disconnected). Such a rewriting exploits the well--known result that any connected graph $\Upsilon $ can be uniquely represented as a tree, whose vertices are 1PI irreducible subgraphs, and whose edges are the bridges of $\Upsilon $. From the remarks above, it follows that we need the effective action at tree--level. This can be immediately obtained from the formal expansion (\ref{Wfeynman})   according to 
\begin{align}
&&\Gamma_{(0)}(\psi)=S_{\psi}-\sum_{\Upsilon \in G_{1PI}}\,
\frac{(a)^{l(\Upsilon )}}{|Aut(\Upsilon )|}\,F_{\Upsilon }\left(S_{\psi}\right) \label{effaction}\\
&& -a\,\ln\,\left(\int\,e^{-\frac{1}{2}S_{\mu\nu}^{\psi}\,\eta^{\mu}\eta^{\nu}}D^{\psi}_g\,[\eta] \right)\notag\;, 
\end{align}
where $G_{1PI}$ is the set of isomorphism classes of 1PI graphs without $J$--vertices, and $|Aut(\Upsilon )|$ denotes the size of the corresponding automorphisms group.\\
\\
To proceed further, we need the expression of $S_{\psi}[\eta;\,\alpha]\doteq S[\exp_{\psi}(\psi_{*}\,\eta);\,\alpha]$. Since we are working in a sufficiently small geodesic ball $B(\phi_{0}, 2r)\subset M$, we can safely assume that  $\psi_{*}\eta$ is small and expand the action  keeping terms only up to second order. Using normal geodesic coordinates in $B(\phi_{0}, 2r)\subset M$, we get after a straightforward computation
\begin{align}
&&S_{\psi}[\eta;\,a^{-1}\,g]=a^{-1}\,\int_{\Sigma}\Big[g_{ij}(\psi)\,\gamma^{\mu\nu}\Big(\partial_\mu\psi^i\partial_\nu\psi^j\notag\\
&&+2\partial_\mu\psi^i\nabla_\nu(\psi_{*}\eta)^j+\nabla_\mu(\psi_{*}\eta)^i\nabla_\nu(\psi_{*}\eta)^j\Big)\notag\\
&&+R_{ijkl}(\psi)\partial_\mu\psi^j\partial_\nu\psi^l(\psi_{*}\eta)^i(\psi_{*}\eta)^k\Big]d\mu_\gamma +O(|\psi_{*}\eta|^3)\nonumber\;,
\end{align}
where $\nabla_\mu(\psi_{*}\eta)^i\doteq\partial_\mu(\psi_{*}\eta)^i+(\psi_{*}\eta)^j\partial_\mu\psi^k\Gamma^i_{jk}(\psi)$ is the pullback of the Levi-Civita connection of $M$ to $\psi^*TM$.\\
\\
Since we have approximated the action $S_{\psi}[\eta;\,a^{-1}\,g]$ to second order in $\psi_{*}\eta$, there will be no vertexes with 3 or more legs in the $\eta$-field theory described by (\ref{effaction}).  This implies in particular that no vacuum 1PI--graphs are possible. Thus (\ref{effaction}) reduces to 
\begin{align}
&&\Gamma_{(0)}(\psi) = a^{-1}\,\int_{\Sigma}g_{ij}(\psi)\,\gamma^{\mu\nu}\partial_\mu\psi^i\partial_\nu\psi^jd\mu_\gamma\label{effzero}\\
\nonumber\\
&&-\ln\Big(\int\exp\Big\{-\frac{1}{2\,a}\int_\Sigma\Big(g_{ij}(\psi)\,\gamma^{\mu\nu}\nabla_\mu(\psi_{*}\eta)^i\nabla_\nu(\psi_{*}\eta)^j\nonumber\\
\nonumber\\
&&+\gamma^{\mu\nu}
R_{ijkl}(\psi)\partial_\mu\psi^j\partial_\nu\psi^l(\psi_{*}\eta)^i(\psi_{*}\eta)^k     \Big)d\mu_\gamma\Big\}D^{\psi}_g[\eta]\Big)\nonumber\;.
\end{align}
The $D^{\psi}_g[\eta]$--integration in $\Gamma_{(0)}(\psi)$ gives rise to a functional determinant which, at face value, is divergent. To make sense of it,  one has expand it in powers of $a$, extract the divergent part to each order and eliminate  it by an opportune redefinition of the metric. 
\subsection{From $\beta$ functions to Ricci flows}
Let $\{e_a\}$ denote a local orthonormal frame in $\psi^{*}TM|_{B(\phi_0,2r)}$, obtained by pulling back an orthonormal frame $\{E_a\}$ defined over $B(\phi_0,2r)$. For notational ease, we shall write $\eta^a$ for the components of $\psi_{*}\eta$ with respect to this $\{e_a\}$. The 
functional integral in (\ref{effzero})  then becomes 
\begin{eqnarray}
\label{gaussin}
\int\exp\Big\{-\frac{1}{2 a}\int_\Sigma(\eta^a\triangle\eta_a+2 (A^\mu)^a_b\eta^b\partial_\mu\eta_a +\\
\nonumber\\
 (A^\mu)^a_b(A_\mu)^c_a\eta^b\eta_c+R_{ajbl}\partial^\mu\psi^j\partial_\mu\psi^l\eta^a\eta^b)d\mu_\gamma\Big\}D^{\psi}_g[\eta]\;, \nonumber
\end{eqnarray}
where we have integrated by parts in the first term, and where $(A_\mu)^a_b$ are the $\{e_a\}$--components of the pullback connection on $\psi^{*}TM|_{B(\phi_0,2r)}$. The Gaussian measure 
$$
D^{\psi}_g[\eta]\,\exp\,-\frac{1}{2 a}\int_\Sigma\eta^a\triangle\eta_a\,d\mu_{\gamma}\,
$$ 
yields a massless field propagator $\langle\eta^a(x)\,\eta^b(y)\rangle$, whereas the remaining terms are treated as interactions. 
The massless field propagator is infrared divergent and needs to be regularized by introducing a small mass term $\mu$. Thus defining  
$\mu:=\Lambda'$, we set 
\begin{equation}
\langle\eta^a(x)\eta^b(y)\rangle = 2\,a\,\frac{\delta^{ab}}{\pi}\int d^2k\frac{e^{ik\cdot(x-y)}}{k^2+\Lambda'^2}
\end{equation}
and let $\Lambda'\to 0$, (\emph{i.e.} $\mu\to 0$), at the end. Expanding (\ref{gaussin}) in Feynman graphs using the above propagator on internal lines and three types of  2-legs vertices: \emph{(i)} $A^\mu\partial_\mu$,\,\, \emph{(ii)} $A^\mu A_\mu$ and \emph{(iii)} ${Rm}\;\partial^\mu\psi\partial_\mu\psi$, we find to 1 loop (\emph{i.e.}, to first order in $a$), three divergent graphs $\Upsilon $, (Fig. 14).
%\begin{figure}[h]
%\begin{center}
%\includegraphics[scale=.8]{fig14}
%\caption{The three graphs contributing to the effective action at 1--loop.}
%\end{center}
%\end{figure}
The first, $\Upsilon_{(i)} $ is a loop with two distinct type \emph{(i)} vertices $x$ and $y$. Its Feynman amplitude is given by
\begin{eqnarray} 
&&F(\Upsilon_{(i)}) = \frac{2}{(a)^2}\int_\Sigma d\mu_\gamma(x)\\
&&\int_\Sigma d\mu_\gamma(y)(A^\mu)^a_b(x)(A^\nu)^c_d(y)\langle\eta^b(x)\partial_\mu\eta_a(y)\rangle\langle\eta^d(y)\partial_\nu\eta_c(x)\rangle\nonumber\;.
\end{eqnarray}
The second graph $\Upsilon_{(ii)}$ is a loop with a type \emph{(ii)} vertex $x$ with amplitude
\begin{equation}
F(\Upsilon_{(ii)})= \frac{1}{2a}\int_\Sigma d\mu_\gamma(x) (A^\mu)^a_b(A_\mu)^c_a\langle\eta^b(x)\eta_c(x)\rangle
\end{equation}
Even though each one of these two amplitudes is divergent, their sum is finite.  Thus, we need to discuss only the amplitude defined by the graph, $\Upsilon_{(iii)}$, associated with a loop with a type \emph{(iii)} vertex $x$,
\begin{equation}
F(\Upsilon_{(iii)})=\frac{1}{2a}\int_\Sigma d\mu_\gamma(x)R_{ajbl}(\psi(x))\partial^\mu\psi^j(x)\partial_\mu\psi^l(x)\langle\eta^a(x)\eta^b(x)\rangle
\end{equation}
We regularize this integral by putting a cutoff $\Lambda$ in the space of momentums. Such a cut--off has a geometrical origin in the fact that we wish to integrate over $\eta$--fields which confine the corresponding $\phi$--fields in the geodesic ball $B(\phi_0,2r)$, \emph{i.e.} we require that $|k|\leq \Lambda$, with $\Lambda ^{-1}<2r$. Thus 
\begin{eqnarray}
F(\Upsilon_{(iii)})&= \frac{1}{2a}\int_\Sigma d\mu_\gamma R_{ajbl}(\psi)\partial^\mu\psi^j\partial_\mu\psi^l\;2\,a\,\frac{\delta^{ab}}{\pi}\int_{|k|\leq\Lambda}d^2k\frac{1}{k^2+\Lambda'^2}=\nonumber\\
\nonumber\\
&=\ln\left(\frac{\Lambda^2}{\Lambda'^2}\right)\int_\Sigma R_{ij}(\psi)\partial^\mu\psi^i\partial_\mu\psi^j d\mu_\gamma
\end{eqnarray}
Then the 1PI effective action (\ref{effzero}) becomes
\begin{eqnarray}
&&\Gamma_{(0)}(\psi) = \int_\Sigma g_{ij}(\psi)\partial^\mu\psi^i\partial_\mu\psi^jd\mu_\gamma +\label{finpart}\\
&&+\,a\,\left(\ln\left(\frac{\Lambda^2}{\Lambda'^2}\right)\int_\Sigma R_{ij}(\psi)\partial^\mu\psi^i\partial_\mu\psi^j d\mu_\gamma +
{finite\quad part}\right)+ \mathcal{O}(\alpha'^2)\;,\nonumber
\end{eqnarray}
where \emph{finite part} indicates terms that are not singular in the limit $\Lambda/\Lambda'\to\infty$. The standard procedure, (minimal subtraction), now consists in regarding the metric $g$ in the first term of (\ref{finpart}) as formally infinite and extracting from it a divergent part so to cancel the 1-loop singularity:
\begin{equation}
g_{ij}(\psi) = g_{ij}(\Lambda/\Lambda') -2\,a\,\ln(\Lambda/\Lambda')\,R_{ij}(\psi) + O(a^2)\;.
\label{renmet}
\end{equation}
The metric $g(\psi)$ in the left hand side is the \emph{bare metric} and $g(\Lambda/\Lambda')$ is the \emph{renormalized metric}. $R_{ij}(\psi)$ is the Ricci tensor of the bare metric, but \emph{we can as well substitute it with that of the renormalized metric}, $R_{ij}(\psi)\Leftrightarrow R_{ij}[g(\Lambda/\Lambda')]$, since the two metrics are equal to order 0 in $\alpha'$. Substituting (\ref{renmet}) into (\ref{finpart}) we finally get
\begin{equation}
\Gamma_{(0)}(\psi)= \int_\Sigma g_{ij}(\Lambda/\Lambda')\,\partial^\mu\psi^i\partial_\mu\psi^j d\mu_\gamma + a\,({finite\,\,\, part}) + \mathcal{O}(a^2)
\end{equation}
Notice that this procedure does not depend explicitly on the point $\phi_0$ in the geodesic neighborhood of which, $B(\phi_0,2r)$,  we are working, \emph{i.e.} the splitting 
(\ref{renmet}) of the bare metric can be extended smoothly to all $M$. Thus, one can extend the above result to the full nonlinear sigma model (that is to background fields $\psi$ taking values in a geodesic neighborhood  $B(\phi_0,2r)$ of any point $\phi_0$).\\
\\
The renormalizability of the theory depends on the behavior of $g(\Lambda/\Lambda')$ when $\Lambda/\Lambda'\to\infty$; this behavior is described by the beta function (\ref{betaflow}), that we can easily compute from (\ref{renmet}). Indeed, by defining $\tau\doteq \ln(\Lambda/\Lambda')$, we immediately get
\begin{equation}
0=\frac{\partial}{\partial\tau}g_{ij} = \frac{\partial}{\partial\tau}\,g_{ij}(\tau)-2a\,R_{ij}(g(\tau)) +\mathcal{O}(a^2)\;.
\end{equation}
Introducing the parameter $t\doteq-a\,\tau$, so that $\partial_tg$ has the same dimension of Ric, one can conclude that the RG flow of the nonlinear sigma model at one loop is \cite{DanThesis}
\begin{equation}
\frac{\partial}{\partial t}\,g(t) = -2{Ric}(g(t))+\mathcal{O}(a^2)\;.
\end{equation}
At this point, it is important to recall that a more detailed analysis  at two loops \cite{DanThesis} would have produced 
\begin{equation}
\frac{\partial }{\partial t}\,g_{ik}(t)=-2\,R_{ik}(t) \,-\,a\,(R_{ilmn}R_{k}^{lmn})+\,\mathcal{O}(a^{2})\;.
\label{2loops}
\end{equation}
Both these RG flow expressions, in the weak coupling limit $a\to 0$, yield  the Ricci flow
(R. Hamilton, '82) \cite{11}
\begin{equation}
\frac{\partial }{\partial t }g_{ab}(t )=-2{R}_{ab}(t )\,,\quad g_{ab}(t =0)=g_{ab}\;. 
\label{rf}
\end{equation}

\subsection{Remarks on the Ricci flow}
Let us recall that the Ricci flow associated with a Riemannian manifold $(M,g)$ can be thought of as the dynamical system on $\mathcal{M}et(M )$ generated by the weakly-parabolic diffusion--reaction PDE \cite{11}
\begin{equation} 
\begin{tabular}{l}
$\frac{\partial }{\partial t }g_{ab}(t )=-2\mathcal{R}_{ab}(t ),$ \\ 
\\ 
$g_{ab}(t =0)=g_{ab}$\, ,\;\; $0\leq t <T_{0}$\;,%
\end{tabular}
   \label{mflow}
\end{equation}
where $\mathcal{R}_{ab}(t )$ is the Ricci tensor of the metric $g_{ik}(t )$. It follows from the above characterization that the geometrical and analytical properties featuring in the Ricci flow are the study of  non--linear parabolic systems of PDEs and the structure theory for Riemannian manifolds. The flow $(M ,g) \mapsto (M ,g(t))$, defined by (\ref{mflow}), always exists in a maximal interval $0\leq t \leq T_{0}$,  for some $T_{0}\leq \infty $, (Hamilton--Shi theorem). This is called a \emph{maximal solution} for the Ricci flow. If such a $T_{0}$ is finite then 
$\lim_{t \nearrow T_{0}}\, [\sup_{x\in M }\,|Rm(x,t )|]=\infty $, \cite{11,12} where $Rm(t)$ is the Riemann tensor of $(M ,g(t ))$, (Fig. 15).  Note that, by exploiting  a  result by N. Sesum and M. Simon\cite{sesum, simon2}, (see also the comments in \cite{knopfsimon})
the curvature singularity regime for the $n$--dimensional Ricci flow is equivalent to $\limsup_{t \nearrow T_{0}}\, [\max_{x\in \Sigma }\,|Ric(x,t )|]=\infty$.
%\begin{figure}[h]
%\begin{center}
%\includegraphics[scale=.9]{fig15}
%\caption{The Ricci flow is generically characterized by the competition between diffusion and concentration of curvature.}
%\end{center}
%\end{figure}
Let $\epsilon \mapsto g^{(\epsilon )}_{ab}(t)$, $0\leq\epsilon \leq 1$, be a smooth one--parameter family of Ricci flows in a
tubular neighborhood $\Omega _{\rho }(g(t))\subset \mathcal{M}et(M)$ of a fiducial Ricci flow $t\mapsto g(t)$.  For $\epsilon \searrow 0$, this set $\{g^{(\epsilon )}_{ab}(t )\}$ is locally characterized by the tangent vector  $h_{ab}(t)$ in $T_{g(t)}\mathcal{M}et(\Sigma )$, covering the fiducial curve $t \rightarrow g_{ab}(t)$, $0\leq t  <T_0$, and defined by the first jet $h_{ab}(t)\doteq \frac{d}{d\epsilon }g^{(\epsilon )}_{ab}(t )|_{\epsilon =0}$ of
 $g^{(\epsilon )}_{ab}(t )$. Any such $h_{ab}(t)$ satisfies the linearized Ricci flow equation
\begin{equation} 
\begin{tabular}{l}
$\frac{\partial }{\partial t }\,h_{ab}(t)=-2\,\frac{d}{d\epsilon }\mathcal{R}^{(\epsilon )}_{ab}(t )|_{\epsilon =0}\,\doteq -2\,D\,\mathcal{R}ic(g(t))\circ \,h_{ab}(t)$\, \\ 
\\ 
$h_{ab}(t =0)=h_{ab}$\, ,\;\; $0\leq t <T_0$\;.%
\end{tabular}
   \label{mflowlin}
\end{equation}
This linearization is not parabolic, due its equivariance under diffeomorphisms. However, there is a natural choice \cite{01}, (see also Chap.2 of \cite{Nibook}), for fixing the action of the diffeomorphism group $\mathcal{D}iff(\Sigma )$, and (\ref{mflowlin}) takes the form of  the parabolic flow $t \mapsto {h}_{ab}(t )\in T_{g(t)}\mathcal{M}et(\Sigma )$ defined, along the fiducial Ricci flow $t \mapsto g(t)$, by  the Lichnerowicz heat equation 
\begin{equation} 
\begin{tabular}{l}
$\bigcirc _{L}\,{h}_{ab}(t  )\doteq \left(\frac{\partial  }{\partial  t   }-\Delta _{L}\right)\,{h}_{ab}(t  )=0\;,$ \\ 
\\ 
${h}_{ab}(t  =0)=\,{h}_{ab}$\, ,\;\; $0\leq t  <T_0$\;,%
\end{tabular}
  \label{divfree}
\end{equation}
where $\Delta _{L}\,:$ $C^{\infty }(\otimes ^{2}T^{*}\,\Sigma)\rightarrow C^{\infty }(\otimes ^{2}T^{*}\,\Sigma)$ is the Lichnerowicz-DeRham Laplacian \cite{lichnerowicz}  on symmetric bilinear
forms defined, (with respect to $g_{ab}(t )$), by 
\begin{equation}
\Delta _{L}{h}_{ab}\doteq \triangle {h}%
_{ab}-R_{as}{h}_{b}^{s}-R_{bs}{h}_{a}^{s}+2R_{asbt}{h%
}^{st},
\label{LDR}
\end{equation}
$\triangle \doteq g^{ab}(t)\,\nabla _{a}\,\nabla _{b}$ denoting the rough  Laplacian. 
For each given $t\in [0,T_{0})$,  $\Delta _{L}$ is a formally $L^{2}$ self--adjoint, not negative semi--definite,  operator of Laplace type \cite{gilkey}. Let us denote by $\left(\circ ,\circ \right)_{L^2(\Sigma ,d\mu_{g(t)})}$ the $g(\beta)$--dependent $L^2$ pairing between $\mathcal{T}_{(\Sigma ,g(t))}\mathcal{M}et(\Sigma )$ and $C^{\infty }(\otimes ^2_ST\Sigma )$, and define \cite{carfback} the the  backward conjugated flow associated with (\ref{divfree}), generated by the operator 
\begin{equation}
\bigcirc^{*} _{L}\doteq -\frac{\partial }{\partial t }-\triangle _{l} +\mathcal{R}\;,
\label{conjflow0}
\end{equation}
thought of as acting on $C^{\infty }(\otimes ^2_ST\Sigma )$. In particular, if $\eta \mapsto g_{ab}(\eta )$ denotes a backward Ricci flow with bounded geometry on $M\times [0,T_0)$, $\eta:=T_0-t$, then we can introduce \cite{carfback} the (backward) heat kernel $\mathcal{H}^{ab}_{a'b'}(y,x;\eta )$  of the corresponding conjugate linearized Ricci operator, \emph{i.e.}, the solution of  
$\bigcirc ^{*}_{L}\, \mathcal{H}^{ab}_{a'b'}(y,x;\eta )=0$, for  $\eta \in (0,T_0)$, \, with $\mathcal{H}^{ab}_{a'b'}(y,x;\eta\searrow 0^{+} )={\delta }^{ab}_{a'b'}(y,x)$, (the bi--tensorial Dirac measure), and where the operator $\bigcirc ^{*}_{L}$ acts on the variables $(x,\eta)$. Similarly, along the (forward)  Ricci flow on $M \times [0,T_0)$ we can consider the $g(t)$--dependent  fundamental solution
 $\mathcal{L}^{ab}_{a'b'}(x,y;t)$ of the linearized Ricci flow, \emph{i.e.}, $\bigcirc_{L}\, \mathcal{L}^{ab}_{a'b'}(x,y;t )=0$, for  $t \in [0,T_0)$, \, with $\mathcal{L}^{ab}_{a'b'}(x,y; t\searrow 0^{+} )={\delta }^{ab}_{a'b'}(x,y)$, and with $\bigcirc_{L}$ acting on $(y,t)$. Note that by $L^2$--duality we have
$\mathcal{K }^{ab}_{a'b'}(y,x;\eta(t=0))\,=\,\mathcal{L }_{a'b'}^{ab}(x,y;t(\eta=0))$. We shall make use of these kernels momentarily\\
This is also the place for mentioning the basic Perelman's no local collapsing theorem \cite{19} according to which given any solution $t\mapsto g(t)$ of the Ricci flow on $M\times [0,T)$, with $M$ compact and $T<\infty $, there exist constants $\kappa>0 $ and $\rho _0>0$ such that for any pair $(x_0,t_0)\in M\times [0,T)$, the metric $g(t)$ is $\kappa $--non collapsed at $(x_0,t_0)$ on scales smaller than $\rho _0$. This means that, if we have $|Rm(x,t )|\leq r^{-2}$ on the geodesic ball neck $B_{t_0}(x_0,r)\times [t_0-r^2,\,t_0]$, for any $0<r<\rho _0$, then we have $Vol_{t_0}\left( B_{t_0}(x_0,r)\right)\geq \kappa \,r^n$. (This implies, by Cheeger's theorem, that the injectivity radius $inj\,(M,x_0,\,g(t_0))$ of $(M,g(t_0))$ at the point $x_0$, is bounded below by $\delta \,r$ for some positive constant $\delta$).  
In Ricci flow theory there is a standard technique, connected to parabolic rescaling, that is used to study what happens as the Ricci flow approaches a singularity. This is known as \emph{point picking} (\emph{e.g.}, \cite{Ni} p.297):
Assume that $t\rightarrow (M ,g(t))$ is a solution to the Ricci flow defined on a maximal time interval $[0,T)$, where $T\leq \infty $, so that $\sup_{M \times [0,T)}\,|Rm|=\infty $ if $T<\infty $. In order to  understand the singularity which is forming as $t\rightarrow T$, one considers a sequence of almost maximum curvature framed marked points $[(x_{i}, O_i),t_i]\in M\times [0,T)$ where $t_{i}\nearrow T$, and $O_{i}$ denotes an orthonormal frame (with respect to  $(M ,g_{i}(t=0))$) at $x_{i}\in M$. From the given solution $t\rightarrow (M ,g(t))$ one constructs a sequence of solutions $t\rightarrow (\widetilde{M} ,  \widetilde{g}_{i}(t), {x}_i)$ defined by $\widetilde{g}_{i}(t)\doteq K_{i}\,g(t_{i}+K_{i}^{-1}\,t)$ for $t\in [-K_{i}\,t_i,\,K_{i}(T-t_i))$, where  $K_{i}\doteq |Rm(x_{i},t_{i})|$. These rescaled solutions are such that the corresponding curvature is bounded, $|\widetilde{Rm}(x_i,t=0)|=1$ and $|\widetilde{Rm}(x,t)|\leq C$, on $M\times [-K_i\,t_i,0]$. Parabolic rescaling  opens the way to the application of Gromov--Hausdorff techniques in Ricci flow theory, and indeed G. Perelman was able exploit his non collapsing result in order to extend  a basic compactness theorem by Hamilton. This theorem, which is central to the application of Ricci flow theory to Thurston geomerization program and, as we shall see, also in QFT, states that the above rescaled solutions $t\rightarrow (\widetilde{M} ,\widetilde{g}_{i}(t), x_i)$ uniformly converge, (in $C^{\infty }$ on compact sets), to a pointed Ricci flow $(\widetilde{M} ,\widetilde{g}(t), \widetilde{x})$, $-\infty <t<\widetilde{T}$, which is a \emph{complete ancient solution}, with bounded curvature,  $\kappa $--non collapsed on all scales. 
The structure of these ancient solutions is strictly connected with the self--similar solutions generated by the action of $\mathcal{D}iff(M )\times \mathbb{R}_{+}$, where $\mathbb{R}_{+}$ acts by scalings. Recall that these are the gradient Ricci solitons ${\mathcal{R}}_{ab}(t )+\nabla _a\nabla _b\,f\,=\, \varepsilon \,{g}_{ab}$, where $f$ is the potential function of the soliton  and where, up to rescaling, we may assume that $\varepsilon =\,-1,\; 0,\; 1$, (respectively yielding for the expanding, steady, and shrinking solitons). A nice and recent analysis of the subtle interplay between ancient solution and Ricci soliton is provided by \cite{BKN, Cao, Ni, NiWallach}. For further details on the geometrical structure of the Ricci flow see the introductory \cite{6} and the monograph \cite{Nibook}.

\subsection{A natural extension: the dilaton coupling} 

By itself the embedding of the Ricci flow into the renormalization group flow for (\ref{sigmamod}) is  quite a remarkable fact. It is not a coincidence, but actually a result pointing to a deeper connection between QFT and Ricci flow theory. To see these further connections   
consider, in place of the harmonic map action (\ref{sigmamod}), the deformed fiducial action given by, (see \S 2.1 and Fig. 1), 
\begin{equation}
\mathcal{S}[\phi;\alpha]=a^{-1}\,\int_\Sigma\gamma^{\mu\nu}\partial_\mu\phi^i\partial_\nu\phi^j\,g_{ij}\,d\mu_\gamma+a^{-1}\,\int_{\Sigma }\,a\,f(\phi)\,\mathcal{K}\,d\mu _{\gamma }\;,
\label{tachaction}
\end{equation}
suggested \cite{Tseytlin1,Tseytlin2} by the dynamics of string theory in a curved background. 
The renormalization group analysis of (\ref{tachaction}), \cite{DanAnnPhys, shore, Tseytlin3, Tseytlin4}  gives rise to a perturbative $\beta$--flow for the coupling fields $\alpha=a^{-1}\,(g,\,a f)$, which at leading order reads
\\
\begin{equation}
\frac{\partial }{\partial \tau}\,g_{ik}(\tau)=2a\,\left(R_{ik}(\tau) \,+2\nabla _i\nabla _k{f}(\tau)\right)\,+\,\mathcal{O}(a^2)\;,
\label{2loopsf}
\end{equation}
\begin{equation}
\frac{\partial }{\partial \tau}\,{f}(\tau)={c_{0}}- 2a\,\left(\frac{1}{2}\,\Delta {f}(\tau)-|\nabla {f}(\tau)|^{2}\right) \,+\,\mathcal{O}(a^2)\;,
\label{2f}
\end{equation}
\\
(note that here is convenient to use the original, unnormalized, scaling variable $\tau=\,-\,\frac{t}{a}$),
where $c_{0}$ is a parameter, the \emph{central charge}, characterizing the theory, and playing in QFT the role of dimensions. For instance, for  bosonic strings  $c_{0}=\frac{\dim M-26}{6}$.\\
\\
A first observation: if we pass to the scaled variable $t:=\,-\,a\tau$, then (\ref{2loopsf}) is the DeTurck \cite{DeTurck} version of Hamilton's Ricci flow deformed by the action of the $t$--dependent diffeomorphism generated by the gradient vector field $2\,g^{ik}\nabla_k\,f(t)$. 
This can be nicely seen in a renormalization group setting if we consider the scaling variable $t$ as a \emph{fictious coordinate time} and describe the kinematics of the flow (\ref{rf})   in the \emph{parabolic spacetime} $M^{n+1}_{Par}\doteq M \times I$,\; $I\doteq [0,T_{0})\subset \mathbb{R}$. We assume that the diffeomorphism 
\begin{equation}
F_{t }:I\times \Sigma \longrightarrow M^{n+1}_{Par};\;\;\; (t ,x)\mapsto i_{t }(x)\;,
\label{Fdiff}
\end{equation} 
of $I\times M$ onto $M^{n+1}_{Par}$, is the identity map, and that $M^{n+1}_{Par}$ carries the product metric ${}^{(n+1)}g_{par}$, so that in the coordinates induced by $F_{t }$  we can write 
\begin{equation}
(F_{t }^{*}\;{}^{(n+1)}g_{par})=g_{ab}(t )dx^{a}\otimes dx^{b}+dt \otimes dt \;.
\label{Fmetr}
\end{equation}
In such a framework, $\frac{\partial }{\partial t }:M \rightarrow TM^{n+1}_{Par}$, can be interpreted as a vector field, transversal (actually, ${}^{(n+1)}g_{par}$--normal) to the leaves $\{M _{t }\}$, describing the renormalization group flow (\ref{rf}) as seen by observers at rest on $M_{t }$. The  evolution of the metric $g(t)$ can be equivalently described by observers in \emph{motion} on $M_{t }$. To this end,  consider a curve of diffeomorphisms $I\ni t\mapsto \varphi (t )\in \mathcal{D}iff(M ) $, (with the initial condition $\varphi ^{i}(x^{a},t =0)=id_{M }$), and define the vector field $X_{\varphi }:M _{t }\rightarrow TM _{t }$, $X_{\varphi }=\frac{\partial }{\partial t }\varphi (t )$, generating $t\mapsto \varphi (t )$. Such a $t $--dependent  $X_{\varphi }$ provides the velocity field of these non--static observers. Thus,
\begin{equation}
\frac{d}{dt}\,F_{t,\varphi  }=\frac{\partial }{\partial t }+X_{\varphi }\,:M _{t }\longrightarrow T\,M^{n+1}_{Par},
\label{Rtime}
\end{equation}
is the space--time vector field covering the diffeomorphism  $F_{t,\varphi  }$ of $I\times M$ onto $(M^{n+1}_{Par}, {}^{(n+1)}g_{par})$, defining space--time coordinates $(t ,y^{i}=\varphi^{i} (t ,x))$ which describe the curve of embeddings $(t , x)\hookrightarrow (t ,\varphi (t ,x))$ of $M _{t }$ in $M^{n+1}_{Par}$. In terms of the coordinates $(t ,y^{i})$ we can write
\begin{equation}
(F_{t,\varphi   }^{*}\;{}^{(n+1)}g_{par})=\breve {g}_{ab}(t  )(dy^{a}+X^{a}_{\varphi }dt )\otimes (dy^{b}+X^{b}_{\varphi }dt )+dt  \otimes dt  \;,
\end{equation}
where the metric $\breve {g}_{ab}(y^{i},t)$ is $t$--propagated according to 
the flow
\begin{equation} 
\begin{tabular}{l}
$\frac{\partial }{\partial t  }\breve {g}_{ab}(t  )=-2\breve {\mathcal{R}}_{ab}(t  )%
-\mathcal{L}_{X_{\varphi }}\breve {g}_{ab}(t ),$ \\ 
\\ 
$\breve {g}_{ab}(t  =0)=g_{ab}$\, ,\;\; $0\leq t  <T_{0}$\;,%
\end{tabular}
  \label{mflowDT}
\end{equation}
where $\mathcal{L}_{X_{\varphi }}$ denotes the Lie derivative along $X_{\varphi }$.
The connection between (\ref{mflowDT}) and (\ref{rf}) is most easily established if we proceed as in the mechanics of continuous media, when shifting from the \emph{body} (Lagrangian) to the  \emph{space} (Eulerian) point of view. To this end, let us introduce the \emph{substantial} derivative $\frac{D}{D t }\doteq \frac{\partial }{\partial t }+\mathcal{L}_{X_{\varphi }}$ associated with the convective action defined by  $X_{\varphi }$ where  $X_{\varphi }^a:=\nabla^a\,\widehat{f}$. Since 
$\frac{D }{D t  }\,\breve {g}_{ab}(t  )=(\varphi ^{*})^{-1}\;\frac{\partial }{\partial t }\left[\varphi ^{*}\,\breve {g}\right]_{ab}$ and $\varphi ^{*}\,\mathcal{R}ic(\breve{g})=\mathcal{R}ic(\varphi ^{*}\,\breve {g})$, $\mathcal{R}(\breve {g})=\mathcal{R}(\varphi ^{*}\,\breve {g})$, it follows from (\ref{mflowDT}) that the pull--back $t \mapsto (\varphi^{*}\,\breve {g})$ of the flow $t \mapsto \breve {g}_{ik}dy^{i}\otimes dy^{k}$, under the action of the $t  $-- dependent diffeomorphism  $x^{a}\mapsto y^{i}=\varphi ^{i}(x^{a},t)$, solves (\ref{rf}). Moreover, under the same construction it is easily checked that the pull-back $\varphi ^*\widehat{f}$ of the dilaton field $\widehat{f}$, constrained by (\ref{DeTf}), satisfies 
\\
\begin{equation}
\frac{\partial }{\partial t}\,\left({\varphi ^*\widehat{f}}\right)(t)=-\Delta \left({\varphi ^*\widehat{f}}\right)(t)+\left|\nabla \left(\varphi ^*\widehat{f}\right)(t)\right|^2-\mathcal{R}(t)\;,
\label{DeTfpulled}
\end{equation} 
\\
Note that both (\ref{DeTf}) and (\ref{DeTfpulled}) are backward--parabolic PDEs.\\ 
\\
A subtler property comes about by noticing that whereas the fiducial action  (\ref{tachaction}) is not conformally invariant, the pertubative QFT it defines is conformally invariant as long as we choose the metric $g$ and the dilaton $f$ couplings  so that the corresponding $\beta$--functions, defined by the right members of (\ref{2loopsf}) and (\ref{2f}), vanish at the given perturbative order. Explicitly, this provides the conditions
\begin{equation}
2a\,\left(R_{ik}(\tau) \,+2\nabla _i\nabla _k{f}(\tau)\right)\,+\,\mathcal{O}(a^2)=0\;,
\label{conf1}
\end{equation}
and
\begin{equation}
-{c_{0}}+ 2a\,\left(\frac{1}{2}\,\Delta {f}(\tau)-|\nabla {f}(\tau)|^{2}\right) \,+\,\mathcal{O}(a^2)=0\;,
\label{conf2}
\end{equation}
which can be also obtained as extremals of the \emph{effective action} functional on (M,g,f) given by
\begin{equation}
\mathcal{F}_{c_0}[g(\tau),\widehat{f}(\tau)]:=\int_{M}\left[2a\left(\mathcal{R}(g)+\,|\nabla\,\widehat{f}|^2\right)-c_{0}\right]\,e^{-\widehat{f}}\,d\mu_{g}\;,
\label{ceffec}
\end{equation}
where we have set $\widehat{f}(\tau):=2\,f(\tau)$. That such a property should hold is suggested by  basic properties of the beta functions, (see \cite{Tseytlin}, this also points to the relevant references), and  follows explicitly  by considering
the  $\tau$--dependent linearization $\mathcal{D}\,\mathcal{F}_{c_0}[{g(\tau)};\widehat{f}(\tau)]\circ \left(\psi ^{ab}(\tau),\phi (\tau)\right)$ of $\mathcal{F}_{c_0}$ in the direction of an arbitrary variation of the fiducial RG flow of the couplings $\tau \mapsto (g_{ab}(\tau),\, \widehat{f}(\tau))$, \emph{i.e.},
\begin{equation}
g^{ab}_{(\epsilon )}(\tau):=g^{ab}(\tau)+\,\epsilon \,\psi ^{ab}(\tau)\;,\;\; g^{ab}_{(\epsilon )}(\tau)\in \mathcal{M}et(M )\;,\forall \epsilon \in [0,1]\;,
\end{equation}
and
\begin{equation}
\widehat{f}_{(\epsilon )}(\tau):=\widehat{f}(\tau)+\,\epsilon\,\phi (\tau)\;.
\end{equation}
\\
A lengthy but otherwise standard computation,  (see \emph{e.g.} \cite{Glickenstein}, Lemma 5.3), provides
\begin{eqnarray}
&&\mathcal{D}\,\mathcal{F}_{c_0}[{g(\tau)};\widehat{f}(\tau)]\circ \left(\psi ^{ab}(\tau),\phi (\tau)\right):=\left.\frac{d}{d\epsilon }\,
\mathcal{F}_{c_0}[{g_{(\epsilon )}(\tau)};\widehat{f}_{(\epsilon )}(\tau)]\right|_{\epsilon =0}=\notag\\
\nonumber\\
&&-2a\,\int_{\Sigma }\,\psi ^{ab}(\tau )\left(\mathcal{R}_{ab}(\tau )+\nabla_a \nabla_b\, \widehat{f}(\tau)\right){\rm
e}^{-\widehat{f}(\tau)}\,d\mu _{{g(\tau)}}\nonumber\\
\nonumber\\
&&+2a\,\int_{\Sigma }\,\left(\frac{\Psi (\tau )}{2}-\phi (\tau) \right)\,\left(2\triangle \widehat{f}(\tau) -|\nabla \widehat{f}(\tau)|^2 +
\mathcal{R}(\tau)\right){\rm
e}^{-\widehat{f}(\tau)}\,d\mu _{{g(\tau)}}\nonumber\\
\nonumber\\
&&-\,c_{0}\int_{\Sigma }\,\left(\frac{\Psi (\tau )}{2}-\phi (\tau) \right)\,{\rm
e}^{-\widehat{f}(\tau)}\,d\mu _{{g(\tau)}}\nonumber
\end{eqnarray}
\\
\noindent where we have set $\Psi (\tau ):=\psi ^{ab}(\tau )\,g_{ab}(\tau)$. It is readily checked that the linearization $\mathcal{D}\,\mathcal{F}_{c_0}[{g(\tau)};\widehat{f}(\tau)]\circ \left(\psi ^{ab}(\tau),\phi (\tau)\right)$ indeed vanishes,  for arbitray variations $\left(\psi ^{ab}(\tau),\phi (\tau)\right)$, when (\ref{conf1}) and (\ref{conf2}) hold at the leading order in $a$.\\
\\
In string theory the effective action (\ref{ceffec}) governs the low energy limit of the couplings  $(g_{ab}(\tau),\, \widehat{f}(\tau))$, here interpreted as \emph{spacetime fields}. Thus, from a physical point of view, the analysis of (\ref{ceffec})  typically aims  to recover the known low--energy structure of spacetime, \emph{i.e.}, Einstein equations and their generalizations, (via a redefinition of the couplings $(g_{ab}(\tau),\, \widehat{f}(\tau))$, this is possible by carrying out the renormalization analysis in the so called \emph{Einstein frame}).
Similarly, the quest for conformal invariance and the lore of the $c$--theorem by A.B. Zamolodchikov \cite{Zamo} features generalization of the action (\ref{ceffec}) that can be interpreted as $c$--functionals. It is never a virtue to stress what could have happened if..., but it is fair to say that these physical motivations did not draw enough attention to the fact that if we evaluate the variation  
$\mathcal{D}\,\mathcal{F}_{c_0}[{g(\tau)};\widehat{f}(\tau)]\circ \left(\psi ^{ab}(\tau),\phi (\tau)\right)$, under the pointwise measure preserving constraint
$D\,e^{-\widehat{f}(\tau)}\,d\mu_{g(\tau)}\circ \left(\psi ^{ab}(\tau),\phi (\tau)\right)=0$, then we immediately get 
\begin{eqnarray}
&&D\,\mathcal{F}_{c_0}[{g(\tau)};\widehat{f}(\tau)]\circ \left(\psi ^{ab}(\tau),\phi (\tau)\right):=\left.\frac{d}{d\epsilon }\,
\mathcal{F}_{c_0}[{g_{(\epsilon )}(\tau)};\widehat{f}_{(\epsilon )}(\tau)]\right|_{\epsilon =0}=\notag\\
\nonumber\\
&&-2a\,\int_{\Sigma }\,\psi ^{ab}(\tau )\left(\mathcal{R}_{ab}(\tau )+\nabla_a \nabla_b\, \widehat{f}(\tau)\right){\rm
e}^{-\widehat{f}(\tau)}\,d\mu _{{g(\tau)}}\nonumber\;.
\end{eqnarray}
This implies that under the constraint $\frac{\partial }{\partial \tau }\,e^{-\widehat{f}(\tau)}\,d\mu_{g(\tau)}=0$, the flow
\begin{equation}
\frac{\partial }{\partial \tau}\,g_{ik}(\tau)=2a\,\left(R_{ik}(\tau) \,+\nabla _i\nabla _k{\widehat{f}}(\tau)\right)\;,
\label{DeT}
\end{equation}
is the gradient flow  of 
\begin{equation}
\mathcal{F}[g(\tau),\widehat{f}(\tau)]:=\int_{M}\left[2a\left(\mathcal{R}(g)+\,|\nabla\,\widehat{f}|^2\right)\right]\,e^{-\widehat{f}}\,d\mu_{g}\;,
\label{Fper}
\end{equation}
with respect to the  $L^2$ metric  defined by $(U,V)_{L^{2}(M )}\doteq \int_{M }g^{il}\,g^{km}\,U_{ik}\,V_{lm}d\mu _{g}$ for  
$U,\,\,V\,\in\,T_{g}\mathcal{M}et(M)$.  In terms of the scaled variable $t=-a\tau$,  the evolution (\ref{DeT}) and  the constraint $\frac{\partial }{\partial \tau }\,e^{-\widehat{f}(\tau)}\,d\mu_{g(\tau)}=0$, (evaluated along (\ref{DeT})), read
\\
\begin{equation}
\frac{\partial }{\partial t}\,g_{ik}(\tau)=-2\,\left(R_{ik}(t) \,+\nabla _i\nabla _k{\widehat{f}}(t)\right)\;,
\label{DeT2}
\end{equation}
and
\begin{equation}
\frac{\partial }{\partial t}\,{\widehat{f}}(t)=-\Delta {\widehat{f}}(t)-\mathcal{R}(t)\;,
\label{DeTf}
\end{equation} 
\\
in which one immediately recognizes the gradient flow version of the Hamilton--DeTurck flow introduced by Perelman \cite{18}. \\
\\
Note that the functional $\mathcal{F}[g(\tau),\widehat{f}(\tau)]$, (see (\ref{Fper})), which we rewrite in terms of the scaled variable $t$ as
\begin{equation}
\mathcal{F}[g(t),\widehat{f}(t)]:=\int_{M}\left(\mathcal{R}(g)+\,|\nabla\,\widehat{f}|^2\right)\,e^{-\widehat{f}}\,d\mu_{g}\;,
\label{FperT}
\end{equation}
is not invariant under generic diffeomorphisms  $\in \mathcal{D}iff(M )$ since it explicitly depends on the dilatonic field $\widehat{f}$. However, it generates a  $\mathcal{D}iff(\Sigma)$--invariant quantity according to \cite{18}
\begin{equation}
\lambda[g]\doteq \inf\left\{\mathcal{F}[{g};\widehat{f}]\;:\;\widehat{f}\in C^{\infty }(\Sigma,R),\,\int_{\Sigma }e^{-\widehat{f}}\,d\mu_{g}=1 \right\}\;
\end{equation}
or equivalently, by setting $u\doteq e^{-\widehat{f}/2}$, as the first eigenvalue
\begin{equation}
\lambda[g]\doteq \inf\left\{\int_{\Sigma }(\mathcal{R}\,u^2+4\,  |\nabla u|^{2})\,d\mu _{{g}}\;,\int_{\Sigma }u^2\,d\mu_{g}=1 \right\}\;,
\end{equation}
of the operator $-4\,\Delta +\mathcal{R}$. In particular, if $u_{(1)}$ is the corresponding (first) eigenfunction, \emph{viz.}, $(-4\,\Delta+\mathcal{R})u_{(1)}=\lambda[g]\,u_{(1)}$, then $\lambda[g]=\mathcal{F}[{g};-2\,\ln\,u_{(1)}]$.
Let $g(\beta)\mapsto\lambda[g(\beta)]$, $\beta\in[0, T_0]$, be the valuation of the functional $\lambda[g]$ on the Ricci flow $\beta\mapsto g_{ab}(\beta)$ on $M\times [0,T_0]$. Note that $\beta\mapsto \lambda[g(\beta)]$ is a continuous function on $[0,T_0]$ and  $\lambda[g(\beta)]:\mathcal{M}et(\Sigma )\rightarrow R$ is  a continuous functional with respect to the $C^2$--topology on $\mathcal{M}et(M )$, (see \emph{e.g.}, chap. 5 of \cite{Glickenstein} ).\\
\\
 The functional $\lambda[g(t)]$ has the remarkable property of being non--decreasing under the Ricci flow \cite{18}. This is a direct offspring of the gradient flow nature of 
(\ref{rf}) and (\ref{DeTf}). Indeed,  along the Ricci flow $t \mapsto g_{ab}(t )$ on $M\times [0,T_0)$, coupled with the parabolic backward evolution (\ref{DeTfpulled}) of $\eta\mapsto f(\eta)$ on $(M,g(\eta)) \times [0,T_0)$,
\ $\eta \doteq T_0-t $,  

\begin{equation} 
\begin{tabular}{l}
$\frac{\partial }{\partial t }g_{ab}(t )=-2\,\mathcal{R}_{ab}(t )\;,$\;\;\;\;$g_{ab}(t =0)=g_{ab}$\, , \\ 
\\ 
$\frac{\partial f(\eta )}{\partial \eta}=\,\triangle_{g(\eta)} f-|\nabla f|_{g(\eta)}^2
+\mathcal{R}(\eta)\;,$\;\;\;$f(\eta =0)=f$\, ,
\end{tabular}
   \label{eqf}
\end{equation}
(for notational ease we set $f(\eta):=(\,{\varphi ^*\widehat{f}}\,)(\eta)$), one computes \cite{18} 
\begin{equation}
\frac{d}{d t} \mathcal{F}[{g(t)};f(t)]= 2\, \int_{\Sigma }\left|\mathcal{R}_{ik}(t )+\nabla_i \nabla_k\, f(t)\right|^2{\rm
e}^{-f(t)}\,d\mu _{{g(t)}}\geq 0\;.
\end{equation}
In particular, $\mathcal{F}$ is stationary on steady gradient solitons flowing along $\nabla f$.\\
\\
The above properties describe the embedding of the Ricci flow into the renormalization group flow as the leading term in a perturbative  weak coupling expansion.
 This raises the natural question of what happens beyond weak coupling. In particular if next-to-leading order contributions to the renormalization group may provide a geometrical flow which generalizes the Ricci flow.

\subsection{Ricci flow singularities and QFT}

The first obvious observation is that the geometrical evolution in the Ricci flow (\ref{rf}) is weakly parabolic only in the infrared regime for the renormalization group flow. This corresponds to $t\to +\infty$, whereas the limit $\Lambda/\Lambda'\to\infty$, driving the renormalization group to a QFT, corresponds to the backward parabolic regime $t\to-\infty$ for the Ricci flow. If we take this at face value, we would associate renormalizable  
nonlinear $\sigma$ model to those Ricci flows which, starting from the bare metric $g$, can be run backwards in time up to $t=-\infty$ without encountering singularities. These are \emph{ancient solutions} of the Ricci flow, the ones which exists on a maximal time interval $-\infty<t<T_0$, where $T_0<\infty$. However, the correspondence that we generate in this way between Ricci flow and QFT is, to some extent, an oversimplified picture. Ancient solutions are certainly necessary for giving meaning to Ricci flow in a renormalization group perspective, however requiring that they arise from a singularity--free Ricci flow is way too restrictive. Actually, the development of Ricci flow singularities may add much more to an emerging geometrical landscape for QFT.\\
\\
To explain this remark, let us assume that along its  evolution the Ricci--flow metric develops somewhere a region of
large curvature. This is the typical behavior, as recalled in  \S 3.2. From a QFT perspective, as the singularity is approached we are no longer in the weak coupling regime for the defining non-linear $\sigma $ model. Field fluctuations cannot be confined to geodesic balls (see \S 3.1), and we cannot any longer assume that such a fluctuations are governed by a large deviation principle around a \emph{background  field}. In short, the correspondence between the renormalization group flow and the Ricci flow seems to badly break down. One may tentatively  insist that the perturbative  expansion of the renormalization group  still has some degree of validity  and argue that in a strong curvature regime terms such as  $a\,(R_{ilmn}R_{k}^{lmn})$ in (\ref{2loops}) may play a significant role.  However, there are strong indications that this is unlikely, since these curvature terms typically enhance the occurrence of singularities rather than taming them. As a rather trivial example, we can look  at how the term $a\,(R_{ilmn}R_{k}^{lmn})$ affects  the volume of a manifold $(M,g(t))$, evolving along the extended flow (\ref{2loops}). A straightforward   computation provides $Vol(M,g(t))=Vol(M,g(0))\,\exp[-\int_0^t\langle \mathcal{R}(s)+\frac{a}{2}|Riem(s)|^2\rangle_{s}\,ds]$, where $\langle \ldots\rangle_{s}$ denotes the average of the enclosed quantity 
over $(M,g(s))$. Thus, the term $a\,(R_{ilmn}R_{k}^{lmn})$ accelerates the volume collapse of the manifold and does not help much. \\
\\
A possible way out of this situation comes from observing that Ricci flow singularities have a well defined structure, and according to the Perelman--Hamilton compactness theorem (cf. \S 3.2) we can associate to a  singularity developing Ricci flow a \emph{complete ancient solution with bounded curvature}, (the singularity model). It is natural to associate this ancient solution and the corresponding QFT to the strong coupling regime of the original model. From a QFT perspective this implies a rather non--trivial  scheme redefinition of the non--linear $\sigma $--model  fields and of the couplings $a^{-1}(g(t), af(t))$.\\ 
Explicitly, one proceeds by exploiting the point--picking technique briefly described at the end of \S 3.2. Thus, let $t\rightarrow (M ,g(t))$ be a Ricci flow on $M\times [0,T)$, representing the leading weak coupling expansion of the renormalization group flow. Let $\{t_i\}$ an increasing sequence of scales and $\{(y_i, O_i)\}$ a corresponding sequence of  framed points, to be interpreted as the background fields $\psi_i$ discussed in \S 3.1, around which the non-linear $\sigma $-model fields fluctuate. Assume that as  $\{t_i\}\nearrow T$ the corresponding curvature in $\{(y_i, O_i)\}$ increases, approaching a high curvature region.
Consider the rescaled sequence of Ricci flows   $t\mapsto (\widetilde{M},\,\widetilde{g}_{i}(t)\doteq K_{i}\,g(t_{i}+K_{i}^{-1}\,t)$ for $t\in [-K_{i}\,t_i,\,K_{i}(T-t_i))$, where  $K_{i}\doteq |Rm(y_{i},t_{i})|$. As recalled in \S 3.2, along these rescaled solutions the corresponding curvatures are bounded on $M\times [-K_i\,t_i,0]$, and we can introduce  the sequence of associated (backward) heat kernels $\{\mathcal{H}^{ab}_{a'b'}(y_i,x;\eta )\}$  of the corresponding conjugate linearized Ricci operators, \emph{i.e.}, the solutions, parametrized by the marked points $\{y_i\}$, of  
$\bigcirc ^{*}_{L}\, \mathcal{H}^{ab}_{a'b'}(y_i,x;\eta )=0$, for  $\eta \in (0,-K_i\,t_i]$, \, with $\mathcal{H}^{ab}_{a'b'}(y_i,x;\eta\searrow 0^{+} )={\delta }^{ab}_{a'b'}(y_i,x)$. Also, these Ricci flows
 are embedded as a weak coupling expansion into the renormalization group flow for the given non--linear $\sigma $--model. In particular, Perelman's no local collapsing theorem, implying a lower bound to the injectivity radius for $(\widetilde{M},\,\widetilde{g}_{i}(t)\doteq K_{i}\,g(t_{i}+K_{i}^{-1}\,t)$, allows us to apply the background field method of \S 3.1 on geodesic balls centered around the framed points $\{(y_i, O_i)\}$. Finally, Perelman--Hamilton compactness theorem allows to extract from the sequence of weak coupled Ricci flows $\{t\mapsto (\widetilde{M},\,\widetilde{g}_{i}(t)\doteq K_{i}\,g(t_{i}+K_{i}^{-1}\,t)\}$ a Ricci flow $(\widetilde{M} ,\widetilde{g}(t), \widetilde{y})$, $-\infty <t<\widetilde{T}$, which is a \emph{complete ancient solution}, with bounded curvature, and which can be naturally associated with a  local $QFT(\widetilde{y})$ perturbatively defined by the following data:\\
 \emph{(i)} A pointed manifold $(\widetilde{M}, \widetilde{y})$, (note that, generally speaking, 
$\widetilde{M}$ is topologically distinct from the original manifold $M$); \\
\emph{(ii)} A complete metric $\widetilde{g}(t)$ with bounded curvature which is $\kappa $--non collapsed on all scales;\\
\emph{(iii)} A backward heat kernel $\widetilde{\mathcal{H}}^{ab}_{a'b'}(\widetilde{y},x;\eta )$ for the conjugate linearized Ricci flow operator $\bigcirc ^{*}_{L}$, based at the marked point $\widetilde{y}$, (the background field), generated by the sequence of heat kernels $\{\mathcal{H}^{ab}_{a'b'}(y_i,x;\eta )\}$;\\
 \emph{(iii)} The action
 \begin{equation}
\mathcal{S}_{\widetilde{y}}[\phi;\alpha]=a^{-1}\,\int_\Sigma\gamma^{\mu\nu}\partial_\mu\phi^i\partial_\nu\phi^j\,\widetilde{g}_{ij}\,d\mu_\gamma+\ldots \;,
\label{renormaction}
\end{equation}
which unambiguosly characterizes an UV asymptotically free field theory, (the $\ldots$ stands for other coupling fields, \emph{e.g.} the dilaton $\widehat{f}$, needed for a full-fledged analysis of the model).\\
\\
Suppose that $QFT(\widetilde{z})=[(\widetilde{M}_2,\widetilde{z}),\widetilde{g}_2(t),\widetilde{\mathcal{H}}^{ab}_{a'b'}(\widetilde{z},x;\eta )]$ is a second QFT defined according to the prescription above and associated with another strong curvature regime developing, along a sequence of framed points $\{(y_i, O_i)\}$, in the Ricci flow associated with the given $(M,g)$. Let $x_{(ik)}$ be points in   $(M, y_i)\times [-K_i(y_i)\,t_i,0]\cap (M, z_k)\times [-K_k(z_k)\,t_k,0]$, for $(i,k)$ sufficiently large, and let $\widetilde{x}$ the corresponding limit point as $(i,k)\to\infty$.  The $L^2$--duality between the heat kernels of the backward and forward linearized Ricci flow implies   
\begin{eqnarray}
\widetilde{\mathcal{H}}^{ab}_{a'b'}(\widetilde{y},\widetilde{x};\eta(t=0))&=&\widetilde{\mathcal{L}}^{ab}_{a'b'}(\widetilde{x},\widetilde{y};t(\eta=0) )\\
\widetilde{\mathcal{H}}^{ab}_{a'b'}(\widetilde{z},\widetilde{x};\eta(t=0))&=&\widetilde{\mathcal{L}}^{ab}_{a'b'}(\widetilde{x},\widetilde{z};t(\eta=0) )\nonumber\;,
\end{eqnarray}
which can be used to relate $QFT(\widetilde{y})$ with $QFT(\widetilde{z})$ in terms of the QFT associated with the background field based at $\widetilde{x}$.  
%\begin{figure}[h]
%\begin{center}
%\includegraphics[scale=.7]{fig16}
%\caption{The QFTs providing two of the local charts defining the QFT landscape associated with the fiducial manifold $(M,g)$.}
%\end{center}
%\end{figure}
 The picture that emerges (Fig.16) from the strategy briefly sketched above is that of a collection of non--trivial QFTs parametrized by the singular set of the Ricci flow over the fiducial target manifold $(M,g)$. Each different QFTs corresponds to possibly different topological manifolds $\widetilde{M}$, and any such a QFT can be considered as a \emph{coordinate patch} of a sort describing the different strong coupling regimes of the theory at the distinct high curvature regions. 
This suggests an imaginative and stimulating geometrical landscape for the QFT associated with non-linear $\sigma $ models, a landscape which in variety and depth is of great interest for both physics and mathematics.

\section*{Acknowledgements}

The author express his gratitude to the organizers of the RISM Conference for the warm ospitality and the friendly atmosphere in Verbania.

\section*{Figure Captions}

{FIGURE 1. The map $\phi$ and the coupling  field $f$.}\\
{FIGURE 2. Spaces of maps $Map(\Sigma,M)$ parametrized by the space of couplings $\mathcal{C}$.}\\
{FIGURE 3. A \emph{coordinatization} of the deformations of a given fiducial action.}\\
{FIGURE 4. Wiener measure on path space over a Riemannian manifold}\\
{FIGURE 5. Correlations over fluctuating surfaces}\\
{FIGURE 6. Averaging fluctuations over length scales $<t$.}\\
{FIGURE 7. Renormalization as a map in the space of actions.}\\
{FIGURE 8. The geometry of the beta function.}\\
{FIGURE 9. The parameter $a$ and the curvature of $|Riem(g)|$  set the scale at which $(\Sigma,\gamma)$ probes the ambient manifold $(M,g)$}\\
{FIGURE 10. The point--like limit: Nearly Gaussian fluctuations near constant maps}\\
{FIGURE 11. In the point--like limit we can define the center of mass $\psi$ of  $N$ indipendent  copies of the mapping $\phi$}\\
{FIGURE 12. The geometrical set up for discussing fluctuations around the background field $\psi$}\\
{FIGURE 13. Nearly Gaussian distribution of the fields $\eta$ in the linear space of maps  $Map(\Sigma,\psi^*TM)$. This distribution generates the effective action for the background field $\psi$.}\\
{FIGURE 14. The three graphs contributing to the effective action at 1--loop.}\\
{FIGURE 15. The Ricci flow is generically characterized by the competition between diffusion and concentration of curvature.}\\
{FIGURE 16. The QFTs providing two of the local charts defining the QFT landscape associated with the fiducial manifold $(M,g)$.}

\end{document}